\documentclass[aps,prb,showpacs,twocolumn,superscriptaddress]{revtex4-1}

\usepackage{graphicx}
\usepackage{braket}
\usepackage{amsmath}
\usepackage{amssymb}
\usepackage{color}
\usepackage{soul}

\newcommand{\vp}{\varphi}

\begin{document}


\title{Localization landscape theory of disorder in semiconductors I: Theory and modeling}


\author{Marcel Filoche}
\email[]{marcel.filoche@polytechnique.edu}
\affiliation{Laboratoire de Physique de la Mati\`ere Condens\'ee, Ecole polytechnique, CNRS, Universit\'e Paris Saclay, 91128 Palaiseau Cedex, France}

\author{Marco Piccardo}
\affiliation{Laboratoire de Physique de la Mati\`ere Condens\'ee, Ecole polytechnique, CNRS, Universit\'e Paris Saclay, 91128 Palaiseau Cedex, France}

\author{Yuh-Renn Wu}
\affiliation{Graduate Institute of Photonics and Optoelectronics and Department of Electrical Engineering, National Taiwan University, Taipei 10617, Taiwan}

\author{Chi-Kang Li}
\affiliation{Graduate Institute of Photonics and Optoelectronics and Department of Electrical Engineering, National Taiwan University, Taipei 10617, Taiwan}

\author{Claude Weisbuch}
\affiliation{Laboratoire de Physique de la Mati\`ere Condens\'ee, Ecole polytechnique, CNRS, Universit\'e Paris Saclay, 91128 Palaiseau Cedex, France}
\affiliation{Materials Department, University of California, Santa Barbara, California 93106, USA}

\author{Svitlana Mayboroda}
\affiliation{School of Mathematics, University of Minnesota, Minneapolis, Minnesota 55455, USA}



\begin{abstract}
We present here a model of carrier distribution and transport in semiconductor alloys accounting for quantum localization effects in disordered materials. This model is based on the recent development of a mathematical theory of quantum localization which introduces for each type of carrier a spatial function called \emph{localization landscape}. These landscapes allow us to predict the localization regions of electron and hole quantum states, their corresponding energies, and the local densities of states. We show how the various outputs of these landscapes can be directly implemented into a drift-diffusion model of carrier transport and into the calculation of absorption/emission transitions. This creates a new computational model which accounts for disorder localization effects while also capturing two major effects of quantum mechanics, namely the reduction of barrier height (tunneling effect), and the raising of energy ground states (quantum confinement effect), without having to solve the Schr\"odinger equation. Finally, this model is applied to several one-dimensional structures such as single quantum wells, ordered and disordered superlattices, or multi-quantum wells, where comparisons with exact Schr\"odinger calculations demonstrate the excellent accuracy of the approximation provided by the landscape theory. 
\end{abstract}


\maketitle

\section{Introduction}

Alloy semiconductors are ubiquitous in many modern semiconductor devices, where the use of heterostructured materials can drastically improve the device performances (see, e.g., Ref.~\citenum{Weisbuch1991}). The main property engineered here is the bandgap difference between the various materials associated in the heterostructures. However, due to the large lattice mismatch between pure compound semiconductors which would lead to highly defective materials if employed as such, one has to associate binary compounds and alloys of binary compounds, leading to ternary alloys, eventually quaternary alloys. The resulting lattice constants are sufficiently close to obtain growth of high-quality materials. One thus retains part of the band-gap discontinuities between the binary compounds which allows one to confine carriers in double heterostructures or quantum wells.\cite{Alferov2001,Kroemer2001} An additional useful property is the modulation of the refractive index, which proved to provide an additional beneficial effect of crucial importance to achieve room-temperature continuous wave lasers, opening the way to optical telecommunication systems. Many other properties are impacted by alloying, some of them not desirable: for instance, the compositional fluctuations induce an additional scattering mechanism for charge carriers which diminishes the carrier mobilities.

GaN-based compounds are among the semiconductors displaying the largest variety of properties of heterostructures and alloys. In addition to the usual features, they show, due to their large ionic composition and crystalline structure (most often wurtzite along the $c$ axis), spontaneous and piezoelectric fields which strongly impact their electrical and optical properties.\cite{Bernardini1997} They also exhibit significant effects of the intrinsic spatial compositional fluctuations of their alloys. The random indium content in InGaN multiple quantum well (MQW) structures can vary locally from 10\% to 23\% within a few nanometers, for an average composition value of 17\%.\cite{Wu2012} These fluctuations can induce a strong modification of carrier spatial distributions, of recombination rates, and of the overall light emission efficiency of the device. They have long been identified as responsible for the short carrier diffusion lengths which in turn lead to high emission efficiencies in spite of the high density of defects still present in the best grown heterostructures.\cite{Nakamura956}

Accounting for carrier localization induced by the local material disorder in semiconductor materials and devices is a daunting task. Usually, it requires solving the Schr\"odinger equation (both for electrons and holes) for random realizations of the disordered potential, and determining the energies and the spatial structure of the localized quantum states. Actually, this problem relates to the famous Anderson localization phenomenon~\cite{Anderson1958} which has triggered an enormous literature, and several of whose aspects still remain puzzling after more than 50 years of research.\cite{Lee1985, Evers2008, Lagendijk2009} From 1983, early evidence of the role played by localization in semiconductors has been observed in conductivity measurements around the metal-insulator transition of three-dimensional (3D) doped charge-uncompensated silicon.\cite{Rosenbaum1983} Theoretical approaches such as the self-consistent scaling theory of Anderson localization have been able to successfully describe this disorder-induced phase transition.\cite{Vollhardt1982} More recently, Anderson localization has been found to play a key role in the transport properties of low dimensional media such as disordered graphene.\cite{Ostrovsky2006} Yet, most of the theories developed to account for Anderson localization rely on statistical averages (through correlation functions) and scaling hypotheses.\cite{Mello1988, Hofstetter1993} Moreover, the fermionic nature of the carriers and the electron-electron and electron-hole interactions (through the electric field) increase the difficulty to reproduce the complex and intricate behaviors of semiconductors. The existing methods for computing transport needs to go down to the atomistic level using techniques such as non-equilibrium Green's functions and coherent potential approximation, at a very high computational cost.\cite{Kalitsov2012} As a consequence, one still lacks a model able to predict efficiently carrier localization and its consequences in realistic disordered semiconductor devices, with the often added complication of multi-layered heterostructures.

One of the most puzzling aspects of Anderson localization is the strong spatial confinement of the one-particle quantum states, attributed to destructive interferences between different propagation pathways in a disordered potential. Recently, a new theory has been proposed, which allows one to accurately predict the localization regions of the carriers, and the density of states (DOS) in the disordered potential created by the fluctuations of material composition, without having to solve the Schr\"odinger equation.\cite{Filoche2012, Arnold2016} This groundbreaking theory is based on a new mathematical tool, the localization landscape~(LL), which is the solution to a Schr\"odinger-type equation with uniform right-hand side.

We present here the implementation of this tool into semiconductor materials, and its use in a semi-classical transport model of semiconductor devices. We show how the formalism of the theory enables us to efficiently predict in semiconductor structures the wave functions and eigen-energies of the confined states, the overlap between electrons and holes, the DOS, and the carrier distribution. This implementation conserves a local formulation, adding only to Poisson and transport equations a different partial differential equation~(PDE). It accelerates the computation time by several orders of magnitude compared to the Schr\"odinger-Poisson-drift-diffusion~(DD) type approach.

The results of this localization theory are further applied to the specific case of nitride semiconductors in two companion papers,  one showing experiments and theory of the Urbach tail of InGaN quantum wells (Piccardo \textit{et al.},\cite{Piccardo2017} hereafter called LL2), the other on the simulation of full light-emitting diode (LED) structures (Li \textit{et al.},\cite{Li2017} hereafter called LL3).

\section{The localization landscape theory}

\begin{figure}
\includegraphics[width=0.45\textwidth]{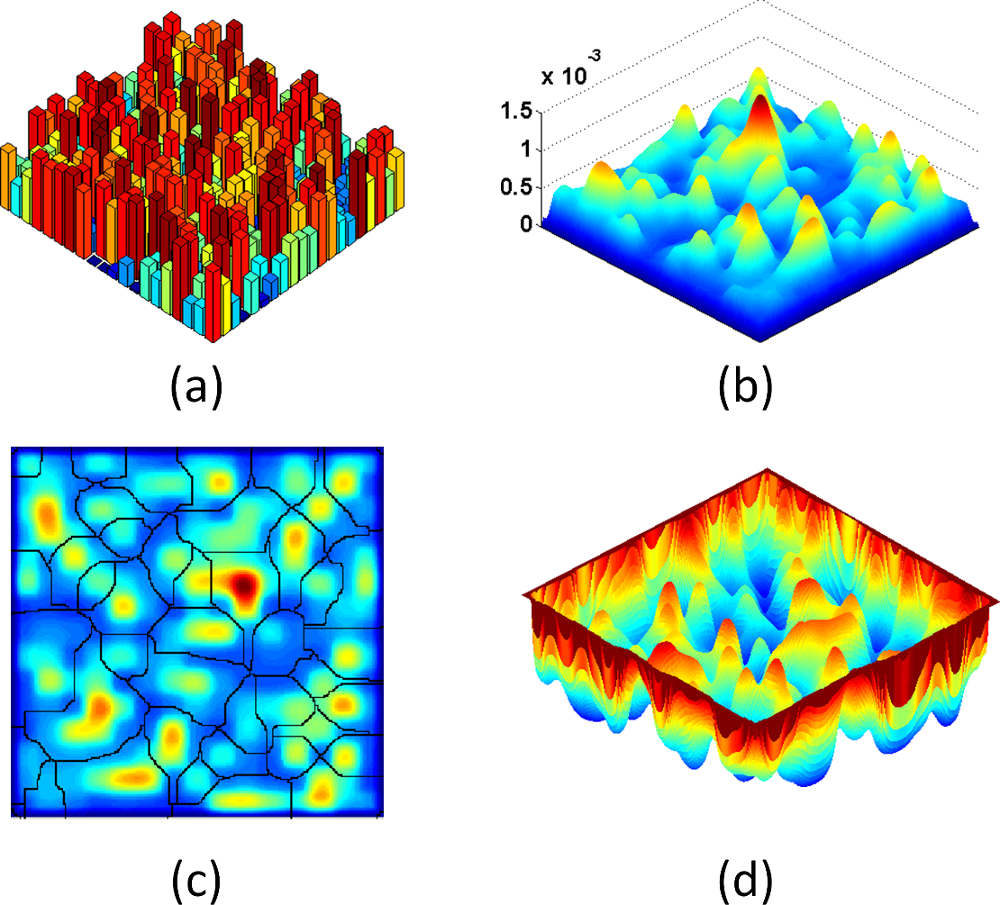}
\caption{The localization landscape theory: (a)~3D representation of the original 2D disordered potential~$V$; (b) 3D representation of the landscape~$u$ solving Eq.~(\ref{eq:udef}). (c)~The valley lines of the landscape~$u$ (black lines) delimit the various localization regions. (d)~Effective localization potential $W \equiv \displaystyle u^{-1}$. The localization subregions outlined in (c) are also the basins of~$W$. \label{fig:landscapes}}
\end{figure}

We present first the main features of the LL theory introduced in Refs.~\citenum{Filoche2012, Arnold2016}. According to this theory, the precise spatial location of  quantum states in a potential $V(\vec{r})$ can be predicted using the solution $u(\vec{r})$ of a simple associated Dirichlet problem, called the localization \emph{landscape}. The quantum states and the energies of particles with mass $m$ are, respectively, the eigenfunctions and the eigenvalues of the Hamiltonian of the system defined as
\begin{equation}
\hat{H} = -\displaystyle \frac{\hbar^2}{2m} \Delta + V~ .
\end{equation}
With this notation, the landscape~$u(\vec{r})$ is defined as the solution to
\begin{equation}
\label{eq:udef}
\hat{H} u = -\frac{\hbar^2}{2m}~\Delta u + V u = 1~ ,
\end{equation}
It is shown in Ref.~\citenum{Filoche2012} that the subregions hosting the localized eigenfunctions are delimited by the valley lines of the graph of~$u$, see Figs.~\ref{fig:landscapes}(a) and (b) for the case of two-dimensional (2D) potential (for a 3D potential, these valleys would be surfaces). This property directly derives from a fundamental inequality satisfied by any eigenfunction~$\psi$ of $\hat{H}$ with eigenvalue~$E$, normalized so that its maximum amplitude is equal to 1 (see Ref.~\citenum{Filoche2012} for the proof):
\begin{equation}
\label{eq:control}
|\psi(\vec{r})| \le E~u(\vec{r})~.
\end{equation}
In other words, the small values of $u(\vec{r})$ along its valley lines~\cite{Filoche2012} constrain the amplitude of~$\psi$ to be small along the same lines and, as a consequence, localize low energy eigenfunctions inside the regions enclosed by these lines. The landscape~$u$ therefore exhibits a partition of the entire domain into a set of subregions, each of these subregions localizing the carriers. But, as exposed hereafter, much more information can be extracted from the localization landscape~$u(\vec{r})$.

\subsection{Effective localization potential}
\label{subsec:effective_potential}

Not only $u(\vec{r})$ controls the eigenfunctions, but also the function $W(\vec{r}) \equiv 1/u(\vec{r})$ (homogeneous to an energy) can be interpreted as a confining potential that is related, among others, to the exponential decay of the Anderson localized states away from their main localization subregion. This property can be proved by transforming the original Schr\"odinger equation through the introduction of an auxiliary function $\psi_1$ such that $\psi \equiv u \psi_1$. A straightforward computation yields:
\begin{equation}
\label{eq:Schrodinger_phi}
-\frac{\hbar^2}{2m} \left[\frac{1}{u^2}~{\rm div}\left(u^2 \nabla \psi_1 \right)\right] + W\psi_1 = E \psi_1~.
\end{equation}
One can see from this equation that the auxiliary function $\psi_1=\psi/u$ thus obeys a Schr\"odinger-type equation in which the original potential~$V(\vec{r})$ has disappeared. Instead, a new function $W(\vec{r})$ now plays the role of an ``effective confining potential''. The valleys of~$u$, which are the boundaries of the localization subregions,\cite{Filoche2012} now correspond to the crest lines (or watershed lines) of~$W$ and thus act as barriers to the auxiliary function~$\psi_1$. This function~$\psi_1$, as well as the initial eigenfunction~$\psi$, is now localized in the basins of $W$ [see Fig.~\ref{fig:landscapes}(c)].

Actually, it was proved that $W$ plays exactly the role of an effective potential thanks to the following identity satisfied by any quantum state~$\ket{\psi}$:
\begin{equation}
\label{eq:W_identity}
\braket{\psi|\hat{H}|\psi} = \frac{\hbar^2}{2m}\braket{u \vec{\nabla} \left(\frac{\psi}{u}\right) | u \vec{\nabla} \left(\frac{\psi}{u}\right) } ~+~ \braket{\psi|\hat{W}|\psi}
\end{equation}
This identity shows that the energy~$E$ of any quantum state~$\ket{\psi}$ can never be smaller than the one it would have in a potential~$W\left(\vec{r}\right)$. Consequently, according to Agmon's inequality,\cite{Agmon1982,Agmon1985} the quantity $(W-E)$ controls exponentially the decay of~$\psi(\vec{r})$ in the regions where $E < W$. Mostly, the eigenfunction decays exponentially with a rate proportional to $\sqrt{W-E}$ in the barriers where $W>E$. More precisely, the decay at point $\vec{r}$ of the amplitude $\psi_{\vec{r}_0}\left(\vec{r}\right)$ of an eigenfunction centered in $\vec{r}_0$ and of energy $E$ is expressed through the inequality:
\begin{equation}
\label{eq:Agmon_ineq}
| \psi_{\vec{r}_0}\left(\vec{r}\right)| \lesssim~e^{\displaystyle -\rho_E(\vec{r_0},\vec{r})}~.
\end{equation}
where $\rho_E(\vec{r_0},\vec{r})$ is the Agmon distance between $\vec{r}_0$ and $\vec{r}$. This Agmon distance (depending on $E$) is defined as:
\begin{equation}
\label{eq:Agmon_dist}
\rho_E\left(\vec{r_0},\vec{r}\right) = \min_{\gamma} \left(\int_\gamma \sqrt{\left( W\left(\vec{r}\right) - E \right)_+}~ds\right)~,
\end{equation}
where $\gamma$ minimum is the geodesic path (i.e. the path of minimum length) between points $\vec{r}_0$ and $\vec{r}$ and $ds$ is the elementary path on that geodesic.

The LL therefore provides an estimate of the decay of the quantum state away from its main existence region. This decay corresponds to the tunneling effect and is more commonly known in quantum mechanics as a result of the Wentzel-Kramers-Brillouin (WKB) approximation. The theory used here is its mathematical generalization, which holds for any potential $W$ satisfying $\braket{\psi|\hat{H}|\psi} \ge \braket{\psi|\hat{W}|\psi}$ for all quantum states. The fact that $W$ plays the role of an effective potential finely shaping the quantum states is crucial for deriving an accurate expression of the density of states, as we will see in Sect.~\ref{sub:3D_DOS}. Finally, our estimate of the eigenfunction amplitudes based on $W$ can also be used, for instance, to assess the coupling between distant localization subregions.\cite{Bondar2011}

\subsection{Eigenvalue and eigenfunction estimates}
\label{subsec:Eigen}

In each of the subregions bounded by the valley lines of the LL~$u(\vec{r})$, the local fundamental eigenfunction and its corresponding energy can also be accurately determined from $u$ itself inside the localization region (the decay far from the localization region being assessed as exposed in Section~\ref{subsec:effective_potential}). To this end, the landscape~$u$, satisfying $\hat{H}u~=~1$, has to be decomposed on the basis formed by the eigenfunctions~$\psi_i$ of the Hamiltonian:
\begin{align}\label{eq:u_decompos}
u &= \sum_i \alpha_i \psi_i\\
 {\rm with} \quad \alpha_i &= \braket{u|\psi_i} = \iiint u(\vec{r})~\psi_i(\vec{r})~d^3r
\end{align}
The decomposition coefficients $\alpha_i$ can be computed using the self-adjointness of the Hamiltonian:
\begin{align}\label{eq:u_coeff}
\alpha_i &= \braket{u|\psi_i} = \frac{1}{E_i} \braket{u| \hat{H} \psi_i} = \frac{1}{E_i}\braket{\hat{H} u |\psi_i} \nonumber\\
&= \frac{1}{E_i} \braket{1|\psi_i} 
\end{align}
From Eqs.~(\ref{eq:u_decompos}) and (\ref{eq:u_coeff}), one can draw three main remarks. First, the lower energy quantum states contribute more to the landscape~$u$ than the high-energy ones (because $E_i$ grows in the denominator of Eq.~(\ref{eq:u_coeff})). Secondly, in a given localization subregion, the low-energy states~($\psi_i$) entering the decomposition of Eq.~(\ref{eq:u_decompos}) are essentially the local quantum states of this subregion. Thirdly, in each subregion, the fundamental state has a bump-like shape, while the higher-energy ones, by orthogonality, take positive and negative values which cancel out so that the scalar products $\braket{1|\psi_i}$ of Eq.~(\ref{eq:u_coeff}) almost vanish. Note that this cancellation also occurs for the high-energy delocalized states of the system. As a consequence, \textit{in each localization subregion} $\Omega_m$, the following relation is deduced: 
\begin{align}\label{eq:u_groundstate1}
u \approx \frac{\braket{1|\psi_0^{(m)}}}{E_0^{(m)}}~\psi_0^{(m)}
\end{align}
$\psi_0^{(m)}$ being the local fundamental state of subregion~$\Omega_m$.
This shows that the local fundamental state $\psi_0^{(m)}$ is almost proportional to $u$ in $\Omega_m$:
\begin{equation}
\label{eq:u_groundstate2}
\psi_0^{(m)} \approx \frac{u}{\Vert u \Vert}
\end{equation}
Inserting Eq.~(\ref{eq:u_groundstate2}) into identity~(\ref{eq:W_identity}) and using the fact that $W\equiv u^{-1}$ allows us to evaluate the fundamental energy $E_0^{(m)}$ from the landscape only:
\begin{align}\label{eq:eigen_energy}
E_0^{(m)} &= \braket{\psi_0^{(m)}|\hat{H}|\psi_0^{(m)}} \approx \frac{\braket{u|\hat{W}|u}}{\Vert u \Vert^2} = \frac{\braket{u|1}}{\Vert u \Vert^2} \nonumber\\
&=\frac{\displaystyle \iiint_{\Omega_m} u(\vec{r})~d^3r}{\displaystyle\iiint_{\Omega_m} u^2(\vec{r})~d^3r} \,
\end{align}
The LL $u(\vec{r})$ therefore provides a direct estimate of the fundamental energy in each of the localization subregions.

\subsection{Density of states}
\label{sub:3D_DOS}

Finally, the prediction of the localized energies extends to the prediction of the integrated density of states (IDOS), hence to its derivative, the density of states (DOS). Thanks to the LL~theory, these quantities can be computed not only globally for the whole system under consideration, but also locally. We detail here the general case of a 3D system as well as the specific case of a 1+2D-system exhibiting two-dimensional translation invariance.

\subsubsection{3D DOS}

Due to the uncertainty principle, $\Delta x \Delta k \approx 2\pi$, each three-dimensional one-particle quantum state spreads in phase space $(\vec{r},\vec{k})$ on a volume of order $(2\pi)^3$. As a consequence, the number of energy states below a given energy $E$ (i.e., the counting function, also called integrated density of states IDOS) is asymptotically equivalent to $\mathcal{V}(E)/(2\pi)^3$ when $E\rightarrow +\infty$, where $\mathcal{V}(E)$ is the volume in phase space determined by $H(\vec{r},\vec{k}) \le E$. This asymptotic behavior is the so-called Weyl's law.\cite{Strauss2008} In a three-dimensional semiconductor, the Hamiltonian of an electron in the conduction band reads as
\begin{equation}
H\left(\vec{r},\vec{k}\right) = \frac{\hbar^2 k^2}{2 m_e^*} + E_c\left(\vec{r}\right)~,
\end{equation}
where $E_c$ is the conduction band energy and $m_e^*$ is the effective electron mass. The IDOS deduced from Weyl's formula is therefore:
\begin{align}
\rm IDOS(E) &= \frac{2}{(2\pi)^3} \iiint_{H(\vec{r},\vec{k}) \le E} d^3r d^3k \nonumber\\
&= \frac{2}{(2\pi)^3} \iiint_{\vec{r}} \left( \iiint_{\frac{\hbar^2 k^2}{2 m_e^*}\le E - E_c(\vec{r})} d^3k \right) d^3r
\end{align}
The factor 2 appearing in the numerator accounts for the spin degeneracy, and the integral within parentheses on the second line is simply the volume of a sphere in $k$-space. The IDOS can therefore be written as a space integral of a local quantity that is classically assimilated to a \emph{local} IDOS, noted LIDOS$(E,\vec{r})$, in the local band structure approximation:
\begin{equation}
\rm IDOS(E) = \iiint_{\vec{r}} \rm LIDOS(E,\vec{r}) ~d^3r\,
\end{equation}
with
\begin{align}\label{eq:N1_E}
\rm LIDOS(E,\vec{r}) &= \frac{2}{(2\pi)^3} \frac{4\pi}{3} \left( \sqrt{\frac{2 m_e^* \left(E - E_c(\vec{r})\right)}{\hbar^2}}\right)^3 \nonumber\\
&= \frac{1}{3\pi^2} \left( \frac{2 m_e^*}{\hbar^2}\right)^\frac{3}{2} \left[E - E_c(\vec{r})\right]^\frac{3}{2}
\end{align}
Differentiating this LIDOS with respect to $E$ gives the local density of states ${\rm LDOS}(E,\vec{r})$:
\begin{align}
{\rm LDOS}(E,\vec{r}) = \frac{1}{2\pi^2} \left( \frac{2 m_e^*}{\hbar^2}\right)^\frac{3}{2} \sqrt{E - E_c(\vec{r})}
\end{align}
One recovers here the classical expression of the local density of states for conduction electrons in a semiconductor.

\begin{figure}
\includegraphics[width=0.4\textwidth]{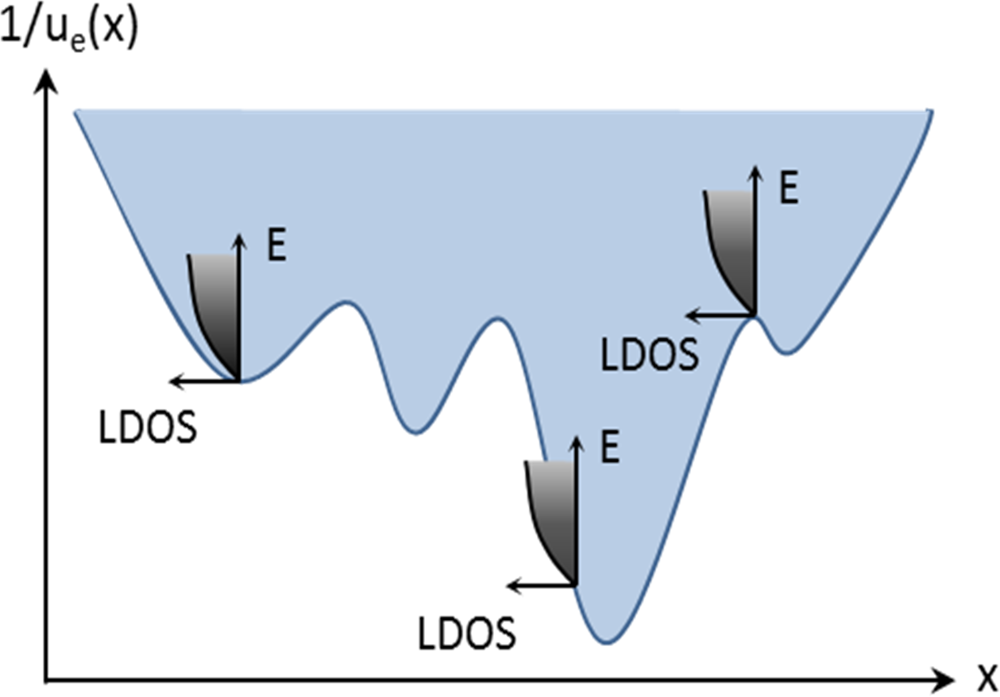}
\caption{Local density of states (LDOS) deduced from the effective potential $W_c \equiv \displaystyle \frac{1}{u_e}$.\label{fig:LDOS}}
\end{figure}

In a disordered system, Eq.~(\ref{eq:W_identity}) shows that $W(\vec{r})$ acts as an effective potential on the particle. Subsequently, $W$ controls the distribution of energies for localized states in each localization subregion. One of the main consequences of this control, as shown on several examples in Ref.~\citenum{Arnold2016}, is that $W$ can be used to accurately estimate the integrated density of states, hence of the density of states also, although the approximate DOS obtained after differentiating the IDOS is in essence less accurate. Practically, this is achieved by replacing the original potential (here $E_c$) by $W$ in the LDOS:
\begin{equation}\label{eq:LDOS3D}
{\rm LDOS}(E,\vec{r}) = \frac{1}{2\pi^2} \left( \frac{2 m_e^*}{\hbar^2}\right)^\frac{3}{2} \sqrt{E - W(\vec{r})}
\end{equation}
This expression physically means that at each point $\vec{r}$, the local density of states is equal to the one of an infinite medium with identical material composition and a parabolic band whose minimal energy would be $W(\vec{r})$ (see Fig.~\ref{fig:LDOS}). In the classical view of the local band approximation, $\vec{r}$ represents in fact a small volume compared to the typical size of the system, but large enough so that it can contain the local electronic states. This approximation which is routinely used when the wave functions are delocalized Bloch waves, is even more justified when dealing with localized eigenfunctions.

The reason for the quality of the approximate IDOS computed using $W_c$ (resp. $W_v$) instead for $E_c$ (resp. $E_v$) has been detailed in Ref.~\citenum{Arnold2016}, but one can give here a short explanation. Weyl's law is fundamentally based on an analogy between quantum and classical filling of phase space, the classical being the volume bounded by $H(\vec{r},\vec{k})=E$, and the quantum originating from the uncertainty principle which states that the volume of a quantum state in phase space is approximately constant (due to $\Delta x . \Delta k \approx 1)$. Thus, counting the number of quantum states of energy smaller than $E$ (the IDOS) comes back to assessing the size of the corresponding volume in phase space. It is known that in a disordered medium, localized states contribute to perturb the bulk IDOS and make it depart from the Weyl's equivalent. Now, the effective potential, which can be seen a smoothed version of the original potential ($E_c$ in the case of electrons), exhibits much better defined wells and barriers. The spatial shapes of the quantum states are closer to the classical trajectories in this effective potential than in the original random or disordered potential. In others words, the effective potential is closer to what is ``experienced'' by the quantum wave if it were a classical particle. As a consequence, the validity of the classical-quantum analogy is strengthened, and the local density deduced from Weyl's law using $W$ much more accurate than any other available estimate.

One needs here to express one word of caution. While IDOS$(E)$ can be understood as the actual number of states below energy $E$ over the spatial region of integration, the ``local'' density of states ${\rm LDOS}(E,\vec{r})$ cannot be understood as the actual number of states at point~$\vec{r}$ as soon as the spatial fluctuations of this LDOS occur on a scale smaller than the typical spatial extension of the electronic states. It should rather be considered as an ``effective'' value, a useful tool for assessing physical quantities.

This remark is of particular importance when this LDOS is integrated over all possible energies to compute the local carrier density $n(\vec{r})$:
\begin{equation}
\label{eq:n_fromLDOS}
n(\vec{r}) = \int_{W(\vec{r})}^{+\infty} \frac{1}{1 + e^{\frac{E-E_F}{k_B T}}}~{\rm LDOS}(E,\vec{r})~dE
\end{equation}
This carrier density is apparently a local quantity at point $\vec{r}$, as in the classical DD model of transport,\cite{vanRoosbroeck1950} but thanks to the effective potential~$W$ appearing in the LDOS, it now also encompasses the quantum confinement induced by the material disorder. As a consequence, this expression is much more accurate than what could be estimated from the classical similar expression with $E_c(\vec{r})$ instead of $W(\vec{r})$. At the same time, it can be easily implemented into a PDE model such as a drift-diffusion solver with Poisson-DD-continuity equations, instead of solving the eigenfunctions of the Hamiltonian.

\subsubsection{1D DOS}

The procedure described above for 3D states can be applied similarly to 1D~systems. In this case, the IDOS reads:
\begin{align}
\label{eq:N_E_1D}
\rm IDOS(E) &= \frac{1}{\pi} \int_{z} \left( \int_{\frac{\hbar^2 k^2}{2 m_e^*}\le E - W(z)} dk \right) dz \nonumber\\
&= \frac{2}{\pi} \int_{z} \sqrt{\frac{2 m_e^*(E-W(z))}{\hbar^2}}~dz\,
\end{align}
and the local density of states is:
\begin{equation}
\label{eq:LDOS_1D}
{\rm LDOS}(E,x) = \frac{1}{\pi} \sqrt{\frac{2 m_e^*}{\hbar^2}} \frac{1}{\sqrt{E-W(z)}}
\end{equation}
If one considers a 3D~system with translational invariance in the two other directions $x$ and $y$, such as a quantum well, then the quantum states are products of 1D and 2D states in the $z$ direction and the $(x,y)$ plane, respectively. The 3D density of states can thus be deduced from the above 1D density of states along $z$ by convoluting it with the 2D LDOS:
\begin{align}
{\rm LDOS}_{3D}(E,z) = \int_{W(z)}^E {\rm LDOS}_{\mbox{2D}}(E-E_1)\nonumber\\
\times \left(\frac{1}{\pi} \sqrt{\frac{2 m_e^*}{\hbar^2}} \frac{1}{\sqrt{E_1-W(z)}}\right)~dE_1\,
\end{align}
LDOS$_{\mbox{2D}}$ being the 2D density of states for free particles, which is constant and equal to $m_e^*/(2\pi\hbar^2)$ (we do not count the spin degeneracy here as it is already included in the 1D~LDOS). This finally gives the 3D~density of states:
\begin{align}
{\rm LDOS}_{\mbox{3D}}(E,z) &= \frac{1}{4\pi^2} \left( \frac{2m_e^*}{\hbar^2}\right)^\frac{3}{2} \int_{W(z)}^E \frac{dE_1}{\sqrt{E_1-W(z)}}\nonumber\\
&= \frac{1}{2\pi^2} \left( \frac{2m_e^*}{\hbar^2}\right)^\frac{3}{2} \sqrt{E-W(z)}
\end{align}
One can notice that the above expression is exactly identical to the 3D~local density of states obtained in Eq.~(\ref{eq:LDOS3D}), except that here $W$ depends only on $z$.

\subsection{Setting the potential reference}

One needs here to underline a peculiarity of the LL. When solving the Schr\"odinger equation, the potential $V(\vec{r})$ experienced by the quantum particle can be defined up to a constant value $K$. If one shifts the potential by~$K$, then the resulting energies are also shifted by the same constant $K$. However, this invariance does not hold for the landscape~$u$. If one considers $u$ being the solution to Eq.~(\ref{eq:udef}), then the solution $u_K$ corresponding to the same potential shifted by a constant $K$ satisfies:
\begin{equation}
-\frac{\hbar^2}{2m} \Delta u_K + (V(\vec{r}) + K)~u_K = 1
\end{equation}
If the constant $K$ is much larger than the typical energies of the quantum states, then $K u_K \approx 1$. Therefore, the corresponding effective potential $W_K = 1/u_K$ is very close to $K$. Inserting the effect of the potential shift on the energies into Eq.~(\ref{eq:control}), the amplitudes of the quantum states are controlled by the landscape through the following inequality:
\begin{equation}
|\psi(\vec{r})| \le (E + K)~u_K(\vec{r})
\end{equation}
In the situation where $K u_K$ becomes very close to 1, this inequality is almost trivially satisfied. In other words, the constraint on $\psi$ exerted by the LL~$u_K$ becomes weaker. This means that the constant $K$ has to be chosen in order to be as small as possible, in such a way that the Hamiltonian remains a positive operator (a condition of applicability of the LL~theory). As a consequence, in semiconductor structures where one encounters large, smooth variations of potential superimposed on small scale random potentials, one should resort to solving the LL piecewise in regions over which the variation of the potential $V$ is of the same order of magnitude than the energies of the quantum states.

\section{The transport model}

\subsection{The new self-consistent approach}

In this section, we present the first implementation of the LL~theory in the physical description of processes in semiconductor heterostructure devices. The exposed model uses an hybrid approach where energy levels, wave functions and DOS are computed using the landscape theory, while a standard description is retained for carrier transport and statistics, as described by the DD~equations of semiconductor textbooks.\cite{Jungel2009} Classically, the DD~model is described through the following set of coupled equations whose unknowns are the electrostatic potential $\vp$ and the quasi-Fermi levels $E_{Fn}$ and $E_{Fp}$:
\begin{equation}
\label{eq:drift-diffusion}
\begin{cases}
{\rm div}\left(\varepsilon_r \vec{\nabla} \vp \right) = \displaystyle \frac{q}{\varepsilon_0} \left(n -p -N_D^+ - N_A^- \pm \rho_{pol} \right)\\
{\rm div}(\vec{J}_n) = R + G_n\\
{\rm div}(\vec{J}_p) = -R + G_p\\
\vec{J}_n = n \mu_n \vec{\nabla} E_{Fn}\\
\vec{J}_p = p \mu_p \vec{\nabla} E_{Fp}\\
n(\vec{r}) = \displaystyle\int_{E_c}^{+\infty} \frac{1}{1 + e^{\frac{E-E_{Fn}}{k_B T}}}~{\rm LDOS}_n(E,\vec{r})~dE\\
p(\vec{r}) = \displaystyle\int_{-\infty}^{E_v} \frac{1}{1 + e^{\frac{E_{Fp}-E}{k_B T}}}~{\rm LDOS}_p(E,\vec{r})~dE
\end{cases}
\end{equation}
where $\varepsilon_r$ is the medium relative permittivity, $n$ and $p$ are the electron and hole densities, $N_A^-$ and $N_D^+$ are the activated doping densities of acceptors and donors; $\rho_{pol}$ is the polarization charge which appears in some semiconductors such as nitrides where electric polarization effects are important; $\vec{J}_n$ and $\vec{J}_p$ are the electron and hole currents, respectively, and $\mu_n$ and $\mu_p$ their mobility; $R$ is the recombination rate which includes all types of recombination processes (SRH, radiative, Auger), and $G_{n,p}$ are the carrier generation rates. Finally, LDOS$_{n,p}$ are the bulk local densities of states for electrons and holes, respectively.

The first equation of (\ref{eq:drift-diffusion}) is the Poisson equation which determines the electrostatic potential from the charge carrier densities. Second and third equations are the continuity equations for both carrier transports. The fourth and fifth equations of (\ref{eq:drift-diffusion}) are the semi-classical expressions of the current densities, derived from the Boltzmann equation assuming a linear collision kernel. These expressions correspond to the linear response theory in statistical physics. The mobilities $\mu_n$ and $\mu_p$ are \emph{effective} parameters that summarize the scattering events and the quantum transport in the bulk materials, either pure compounds or alloys. Finally, the sixth and seventh equations of (\ref{eq:drift-diffusion}) compute the carrier densities from the quasi-Fermi levels $E_{Fn}$ and $E_{Fp}$ through the densities of states LDOS$_{n,p}$.

\begin{figure}
\includegraphics[width=0.45\textwidth]{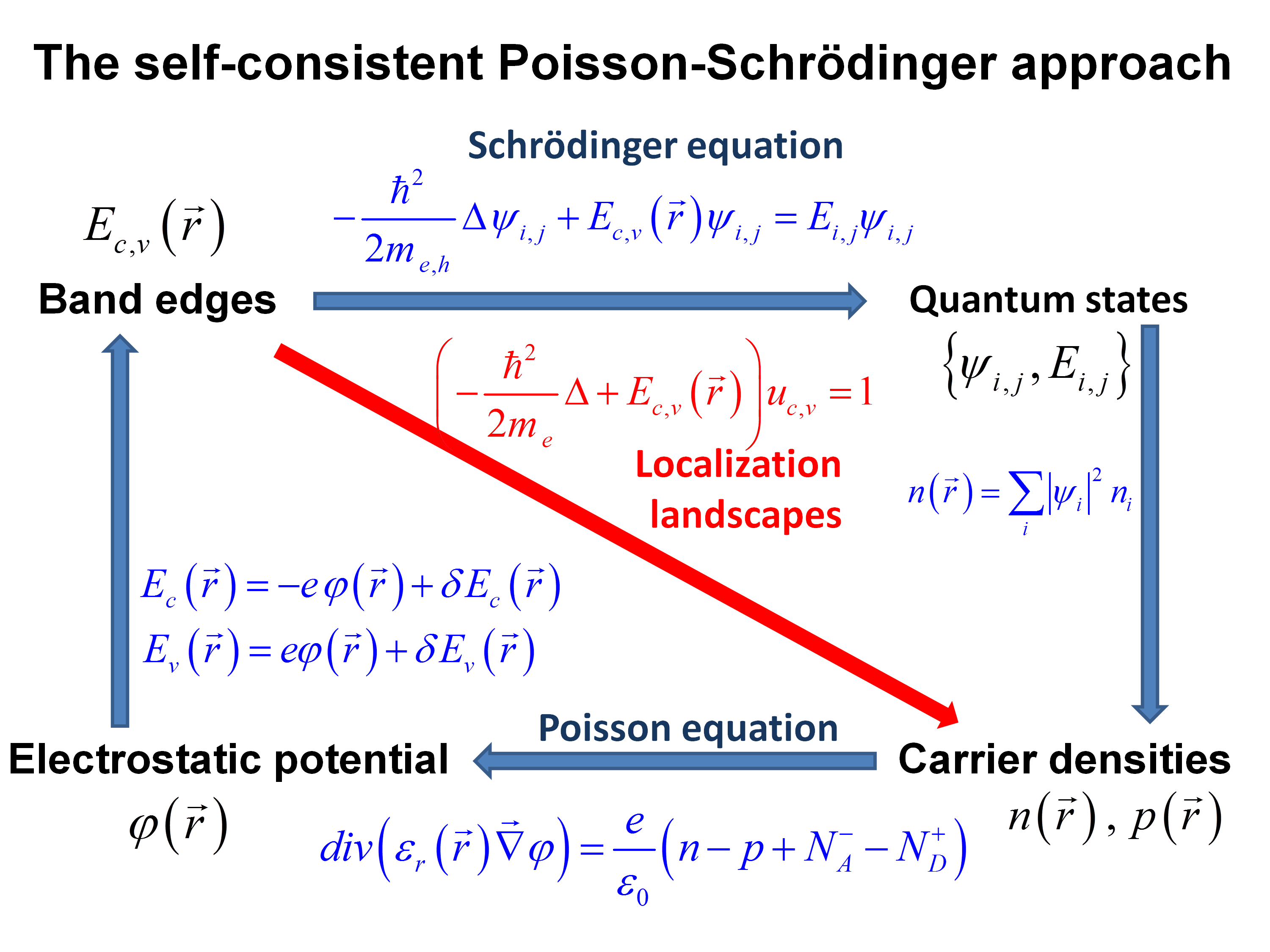}
\caption{Schematic structure of the new self-consistent Poisson-landscape model allowing us to bypass solving the Schr\"odinger equation.
\label{fig:self_consistent}}
\end{figure}

In the usual DD~model, Poisson and transport equations are solved self-consistently to obtain the converged electrostatic potential and the quasi-Fermi levels $E_{Fn}$ and $E_{Fp}$. More advanced models, called hydrodynamic, include also energy transport and hot carriers,\cite{Jungel2009} but at the expense of much larger computational time. In order to account for quantum confinement effects, one has to solve the Schr\"odinger equation using $E_c(\vec{r})$ and $E_v(\vec{r})$ as potentials for electrons and holes, respectively. The carrier densities are then deduced from the electron and hole wave functions, $\{\psi_{e,i}\}$ and $\{\psi_{h,j}\}$, and their corresponding energies, $\{E_{e,i}\}$ and $\{E_{h,j}\}$ through summation with Fermi-Dirac distribution:
\begin{equation}
\begin{cases}
n(\vec{r}) = \displaystyle \sum_i \left(\frac{1}{1 + e^{\frac{E_{e,i}-E_{Fn}}{k_B T}}}\right)~|\psi_{e,i}(\vec{r})|^2\\
p(\vec{r}) = \displaystyle\sum_{j} \left(\frac{1}{1 + e^{\frac{E_{Fp}-E_{h,j}}{k_B T}}}\right)~|\psi_{h,j}(\vec{r})|^2
\end{cases}
\end{equation}
In the absence of currents, these carrier densities enter Poisson equation which in turn modifies the electrostatic potential, and then $E_c$ and $E_v$ (the band offsets $\delta E_{c,v}$ mentioned in Fig.~\ref{fig:self_consistent} being local properties of the material). This is the usual (without disorder) self-consistent Poisson-Schr\"odinger scheme, depicted by the blue cycle in Fig.~\ref{fig:self_consistent}. Accounting for quantum transport adds another level of complexity, which consists in solving the Schr\"odinger equation to determine the equilibrium quantum states, and then compute transitions between these states.\cite{WatsonParris2011} The dynamic carrier densities now enter Poisson and transport equations, from which one deduces the electrostatic potential and the quasi-Fermi levels. This loop results in lengthy, time-consuming and possibly unstable simulations that can take days of computation for a complex 3D structure, even damagingly longer with compositional disorder. At an even more fundamental level, non-equilibrium Green's function techniques or atomistic tight-binding models are used to assess fast processes.\cite{AufderMaur2015, Schulz2015}

Overall, DD-based models still remain to this date the only models able to simulate large structures and compute carrier transport in realistic semiconductor devices with a computational time compatible with optimization and design (see LL3, Ref.~\citenum{Li2017}).

\begin{figure}
\includegraphics[width=0.4\textwidth]{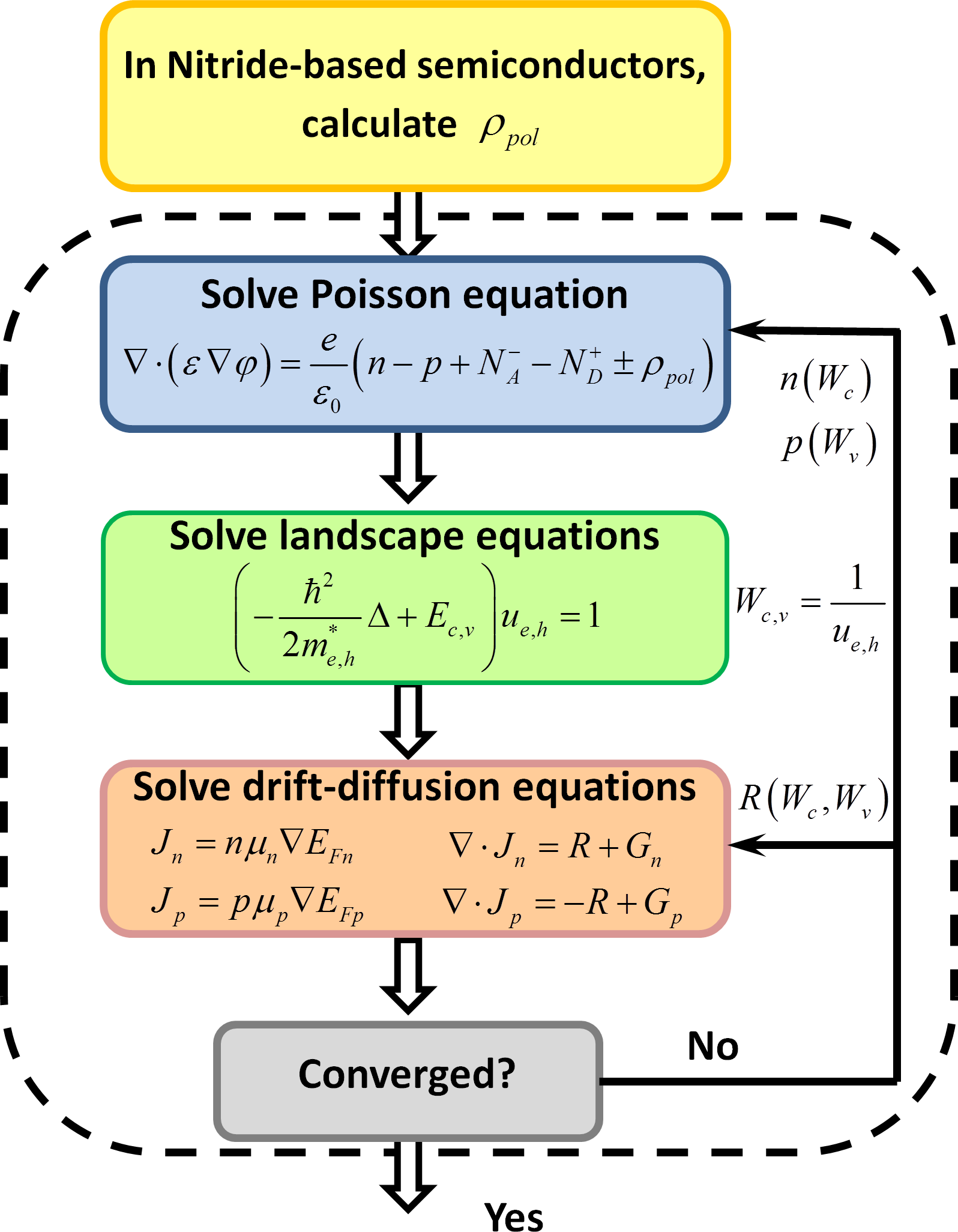}
\caption{Flow chart solving the Poisson and drift-diffusion equations by applying the localization landscape theory.\label{fig:flowchart}}
\end{figure}

In the following, the implementation of the LL~theory is first presented for the Poisson scheme, then for the full DD~model including carrier transport. Introducing the LL~$u(\vec{r})$ allows us to entirely bypass solving the Schr\"odinger equation, a computationally highly demanding step. From the electrostatic potential, we deduce two LLs~$u_e$ and $u_h$, hence, two effective potentials $W_c$ and $W_v$. According to the theory, these landscapes provide direct estimates of the density of states LDOS$_n(\vec{r})$ and LDOS$_p(\vec{r})$ [see Eq.~(\ref{eq:LDOS_1D})], and of the carrier densities. This bypass is depicted by the red arrow in Fig.~\ref{fig:self_consistent}. The recombination rates are also assessed by computing the overlap between electron and hole states in the localization subregions (see next subsection). These quantities are then used to solve Poisson and DD equations. This makes the LL~theory easily compatible with a classical DD~approach. The schematic flow chart of the entire self-consistent simulation process is displayed in Fig.~\ref{fig:flowchart}.

One has to underline that in this model, the current densities still take their classical DD~form expressed in Fig.~\ref{fig:flowchart}. In this form, the tunneling current caused by the spatial tails of the eigenstates that cross potential barriers is inherently accounted for in the model through the lowering of the effective potential $W_c$ (resp. $W_v$) as compared to $E_c$ (resp. $E_v$). However, the quantum current originating from phonon-assisted hopping between different eigenstates or by scattering events, is still accounted for through effective transport parameters, i.e., the mobilities $\mu_n$ and $\mu_p$. Extension of the theory to account for a full quantum model of transport is currently work in progress.

\subsection{Computing generation/recombination processes}

The final subsection of the modeling section is dedicated to the computation of the generation and recombination processes in direct gap semiconductors using the LL, the focus being in particular on optical interband transitions, namely absorption and emission. Typically, the computation of these processes in quantum-confined semiconductor structures requires the knowledge of the eigen-energies and eigenfunctions of the quantum states, which determine respectively the energy of the transitions and the overlap between the electrons and holes. As derived in Sec.~\ref{subsec:Eigen}, these quantities can be in fact directly assessed from the~LL. It should be noted that due to the integrals appearing in Eq.~(\ref{eq:u_groundstate2}) and~(\ref{eq:eigen_energy}), these relations cannot be implemented into a local formulation, such as a self-consistent Poisson-DD-landscape loop, but can be evaluated after convergence is reached for the simulated system. 

We consider in the following the case of a \textit{homogeneous} quantum well (QW), for which the classical absorption/emission expressions are rewritten using the LL~theory. The extension to the case of QWs with in-plane disorder is briefly outlined, while for more details the reader is referred to the companion papers LL2 and LL3.

\subsubsection{Absorption}
\label{subsubabs}

The quantum mechanical expression of the absorption coefficient $\alpha$ of a homogeneous QW of thickness $L$ as a function of photon energy $h\nu$ is~\cite{Singh2007}
\begin{equation}
\label{eq:classical_absorptionQW}
\alpha(hv) = \frac{C}{L}\cdot g_{2D} \sum_{i,j} \Theta(h\nu-E_g-E_{e,i}-E_{h,j}) ~I_{i,j}\,
\end{equation}
where the prefactor $C=\pi e^2\hbar |p_{cv}|^2/ m_0^2 c n_r \varepsilon_0 h\nu$ depends on the real part of surrounding refractive index $n_r$ and the interband momentum matrix element $p_{cv}$. Considering a narrow photon energy range (close to the band gap), we neglect the $k$-dependence of $p_{cv}$ and the $h\nu$ dependence of $C$, and treat this prefactor as a constant; $g_{2D}=m_r/\pi\hbar^2$ is the joint density of states (JDOS) of a 2-D system including the spin degeneracy with $m_r=(1/m_e^*+1/m_h^*)^{-1}$ being the reduced effective mass. The summation is performed over all electron and hole states (labeled $i$ and $j$, with $i=j=0$ being the fundamental states) with eigen-energies $E_{e,i}$ and $E_{h,j}$. The overlap factor~$I_{i,j}$ is determined by the electron and hole eigenstates $\psi_{e,i}$ and $\psi_{h,j}$ as
\begin{equation}
\label{eq:overlap_psi}
I_{i,j}=\frac{|\braket{\psi_{e,i}|\psi_{h,j}}|^2}{\|\psi_{e,i}\|^2 \| \psi_{h,j} \|^2}
\end{equation}

Our approach here is to use the LL~theory to assess the absorption coefficient directly from the maps of $u_e$ and $u_h$ computed for electrons and holes using Eq.~(\ref{eq:udef}), where $V$ corresponds to the conduction and valence band potential, respectively. The valleys of the landscapes (which are also the crest lines of the $W_{c,v}$ potentials) partition the domain into localization subregions. Note that in the case of a homogeneous QW, only one localization subregion exists for each type of carrier, determined by the confinement along the growth direction, while in the case of a disordered QW several subregions can be found.\cite{Piccardo2017, Li2017} The determination of the valley lines in a homogeneous QW is discussed in Sec.~\ref{sub:eigenen_overlap}.

In each subregion the eigenfunctions of the fundamental states, $\psi_{e,0}$ and $\psi_{h,0}$, and the corresponding eigen-energies, $E_{e,0}$ and $E_{h,0}$, can be calculated using Eq.~(\ref{eq:u_groundstate2}) and~(\ref{eq:eigen_energy}). From the estimation of the eigenstates, the overlap factor $I_{0,0}$ of Eq.~(\ref{eq:overlap_psi}) can then be directly computed. To sum over all interband transitions, Weyl's law can be used to estimate the system DOS instead of summing over all possible $i,j$ as in Eq.~(\ref{eq:classical_absorptionQW}). In Sec.~\ref{sub:3D_DOS} it was shown that in a 3D system exhibiting confinement along one direction and translational invariance along the others, Weyl's law predicts an LDOS exactly identical to that of the bulk case. Although the values of $E_{e,0}$ and $E_{h,0}$ account for the quantum confinement of the carriers in the disordered potential, bulk asymptotic Weyl's law is only a continuous yet good approximation of the discrete energy spectrum. Future works should allow us to provide a discrete and even better estimate of the spectrum in each localization subregion, based on $W$. To compute the JDOS of the homogeneous QW, the following well-known expression for bulk is thus used considering an effective band gap $E_g+E_{e,0}+E_{h,0}$:
\begin{equation}
{\rm JDOS}_{\mbox{3D}}(h\nu)=\frac{\sqrt{2}~m_r^{\frac{3}{2}}}{\pi^2\hbar^3}\sqrt{h\nu-E_g-E_{e,0}-E_{h,0}}
\end{equation}
Finally the absorption coefficient of the homogeneous QW can be rewritten as
\begin{equation}
\label{eq:absorption_QW_landscape}
\alpha(hv) = \frac{2}{3}\cdot C\cdot {\rm JDOS}_{\mbox{3D}}(hv)~I_{0,0}
\end{equation}
where all quantities can be derived from the landscapes without solving the Schr\"odinger equation. (The factor 2/3 in Eq.~(\ref{eq:absorption_QW_landscape}) and the absence of the $1/L$ prefactor appearing in Eq.~(\ref{eq:classical_absorptionQW}) are due to the use of the bulk expression of the JDOS in the absorption coefficient.\cite{Singh2007})

\begin{figure}
\includegraphics[width=0.43\textwidth]{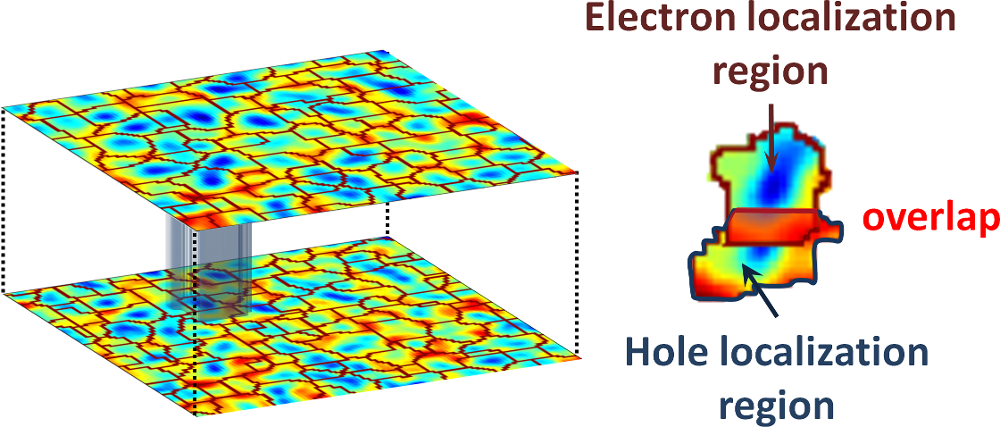}
\caption{Overlap regions are defined as intersections between electron and hole localization subregions. (Left) superimposed electron and hole landscapes. (Right) Example of one overlapping region defined as the intersection between two electron and hole localization subregions.
\label{fig:overlap}}
\end{figure}

The procedure for determining the absorption coefficient in the case of a disordered QW remains essentially the same. When superimposing the maps of the two landscapes, one can define the subregions which are the intersections between the various localization subregions for electrons and holes (Fig.~\ref{fig:overlap}). For each of these electron-hole ``overlapping'' subregions a local value of $\alpha$ can be computed using Eq.~(\ref{eq:absorption_QW_landscape}). The overall absorption coefficient of the disordered QW is obtained by summing over all electron and hole subregions, $\Omega_m$ and $\Omega_n$, as
\begin{equation}
\label{eq:absorption_disorderedQW_landscape}
\alpha(hv) =  \frac{2}{3}\cdot C \cdot \sum_{m,n} {\rm JDOS}_{3D}^{(m,n)}(hv)~I_{0,0}^{(m,n)}~,
\end{equation}
where the JDOS and overlap factors depend on the fundamental state and corresponding energy of the considered subregion, namely $\psi_{e,0}^{(m)}$ and $E_{e,0}^{(m)}$ for electrons, and $\psi_{h,0}^{(n)}$ and $E_{h,0}^{(n)}$ for holes, which again can be calculated from the landscapes using Eq.~(\ref{eq:u_groundstate2}) and (\ref{eq:eigen_energy}). Such an accurate accounting of carrier localization in a disordered material is crucial for a precise assessment of the below-gap absorption processes. A detailed study on this topic is presented in Ref.~\citenum{Piccardo2017}.

\subsubsection{Emission}

Let us first discuss the homogeneous case. While in the previous derivation of optical absorption [Eq.~(\ref{eq:absorption_QW_landscape})] the conduction and valence bands were considered as completely empty and filled, respectively, in the case of optical emission the occupation of the states depends on the specific injection conditions of the QW. Assuming that the quasi-Fermi levels for electrons and holes are known, the electron and hole distributions $n(z)$ and $p(z)$ in the QW can be determined using Eq.~(\ref{eq:n_fromLDOS}). The radiative recombination rate then reads as
\begin{equation}
R_{int} = \int B_0(z)~n(z)~p(z)~dz
\end{equation}
corresponding to the transition energy $E_g+E_{e,0}+E_{h,0}$ calculated from the landscapes as in Sec.~\ref{subsubabs}. $B_0(z)$ is the local radiative recombination coefficient. If the transition is homogeneously broadened then a Lorentzian function must be used to determine the emission spectrum of the non disordered QW as
\begin{equation}
R_{sp}(h\nu) = \frac{\displaystyle \frac{\Gamma}{2\pi}~R_{int} }{\left(h\nu-E_g-E_{e,0}-E_{h,0}\right)^2 + \left(\displaystyle \frac{\Gamma}{2}\right)^2}
\end{equation}
In the case of a disordered QW, the emission spectra have to be summed over all possible transitions between localized states to produce the inhomogeneously broadened luminescence spectrum of the QW (see companion paper LL3, Ref.~\citenum{Li2017}).

The localization landscapes can also be used to compute non-radiative recombination processes which may be strongly affected by disorder. In Auger recombination the confinement increases the carrier momentum compared to free carriers, leading to a better overlap with the wave function of the final high-energy carrier, and thus an enhanced transition probability. Let us consider for instance the ``hhe'' Auger process: an electron and a hole recombine and transfer their energy through Coulomb interaction to a second hole which becomes highly energetic (hot). Using the Fermi golden rule, one expresses the Auger recombination rate in one overlapping region:
\begin{equation}
\frac{1}{\tau} = \frac{2\pi}{\hbar}\lvert M_{if}\rvert^2\rho(E_f)\,
\end{equation}
where $\rho(E_f)$ is the density of final states and $M_{if}$ is the matrix element defined as:
\begin{align}\label{eq:Mif}
M_{if} &= \sqrt{2}\iint dr_1 dr_2~{\psi_{h,0}}^*(r_1){\psi_{h,0}}^*(r_2) \nonumber\\ & \times V(r_1,r_2)~\psi_{e,0}(r_1)~\varphi_{h,f}(r_2)\,
\end{align}
$r_i$ being the initial and final positions, and $V$ the Coulomb interaction potential. $\psi_{h,0}(r_1)$ and $\psi_{h,0}(r_2)$ are the initial states of the two holes, while $\psi_{e,0}(r_1)$ is the initial electron state. Finally, $\varphi_{h,f}(r_2)$ is the final state of the second hole.
After dividing the entire system into the aforementioned overlapping regions, the above integral is computed over each region separately. For instance, in the above integral, the final hole wave function $\varphi_{h,f}$ is assumed to be a simple plane wave of wave vector $k_f$, while the initial states are approximated by the local landscapes in the considered overlapping region (see Section~\ref{subsec:Eigen}). All these local integrals can then be assembled to build maps of the Auger recombination times in the whole system. The influence of compositional disorder on Auger recombination tackled by the localization landscape theory is currently under study.

\section{Applications of the landscape theory to simple 1D heterostructures}

In the following sections, we study several structures to show in various configurations the ability of our model to capture the main features of quantum devices. We successively study the eigen-energies and the overlap between quantum states in single QW (SQW) and 3-QW~structures~(Section~\ref{sub:eigenen_overlap}). We then apply our model to the computation of density of electronic states in 1-D periodic and disordered superlattices (Section~\ref{sub:SL_DOS}). Finally in Section~\ref{sub:carrier_comparison}, a 1-D SQW~structure is simulated to perform a detailed comparison of the carrier distributions computed by the classical Poisson equation, the Poisson-landscape model (LL~theory), and the Poisson-Schr\"odinger equation, respectively. All modeled structures presented in the following are 1-D with homogeneous layers. A full 3-D modeling of carrier transport in a semiconductor device including the effect of compositional disorder is presented in the companion paper LL3 (Ref.~\citenum{Li2017}).

\subsection{Eigen-energies and overlap}
\label{sub:eigenen_overlap}

In GaN-based materials, the polarization field at hetero-interfaces can strongly depend on the growth direction due to the wurtzite crystal structure. Conventionally, GaN grown along the (0001) direction ($c$-plane) cannot avoid a lattice mismatch which in turn induces a piezoelectric polarization. In addition, the relative displacement of the cation and anion sub-lattices from the ideal wurtzite position generates a net spontaneous polarization. In contrast, GaN grown along the (1$\bar{1}$00) direction ($m$-plane) is non-polar and does not exhibit any strain-induced piezoelectric polarization or spontaneous polarization field. In the simulation framework developed here, the density of polarization charges~$\rho_{pol}$ is computed by taking the divergence of the total polarization ($\nabla\cdot$P$^{\rm total}$), including spontaneous and piezoelectric polarization fields to account for the internal electric field at the hetero interfaces. Spontaneous and piezoelectric polarizations are computed using Eq.~(\ref{eq:psp}) and (\ref{eq:ppz}), while parameters of polarization values and piezoelectric coefficients can be found in Table~\ref{tab:parameter-pol} and \ref{tab:parameter-piezo}, respectively (see Appendix).

\begin{figure}
\includegraphics[width=0.45\textwidth]{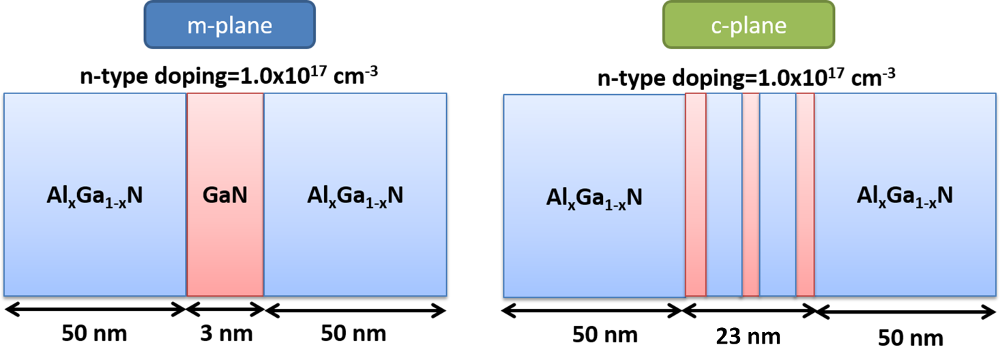}
\caption{QW~structures used to test the LL~theory. (Left) SQW well with $m$-plane orientation of the well material. (Right) 3-QW structure with $c$-plane orientation. The well width and the barrier thickness are 3 and 7~nm, respectively.
\label{fig:MQW_structures}}
\end{figure}

First, we test the ability of the Poisson-LL model to accurately predict the energies and the spatial extension of quantum states in QW~structures. Two different types of structures are simulated (Fig.~\ref{fig:MQW_structures}). The first one is a 3~nm~SQW of $m$-plane GaN enclosed between 50~nm barriers of Al$_x$Ga$_{1-x}$N, several values of Al fraction $x$ being computed in a range from 0.1 to 0.5. The second structure is formed of 3~QWs of $c$-plane GaN of 3~nm width separated by 7~nm barriers of Al$_x$Ga$_{1-x}$N, also enclosed between 50~nm barriers of Al$_x$Ga$_{1-x}$N. This allows us to test the confinement effects in the wells in our model.

\begin{figure}
\includegraphics[width=0.23\textwidth]{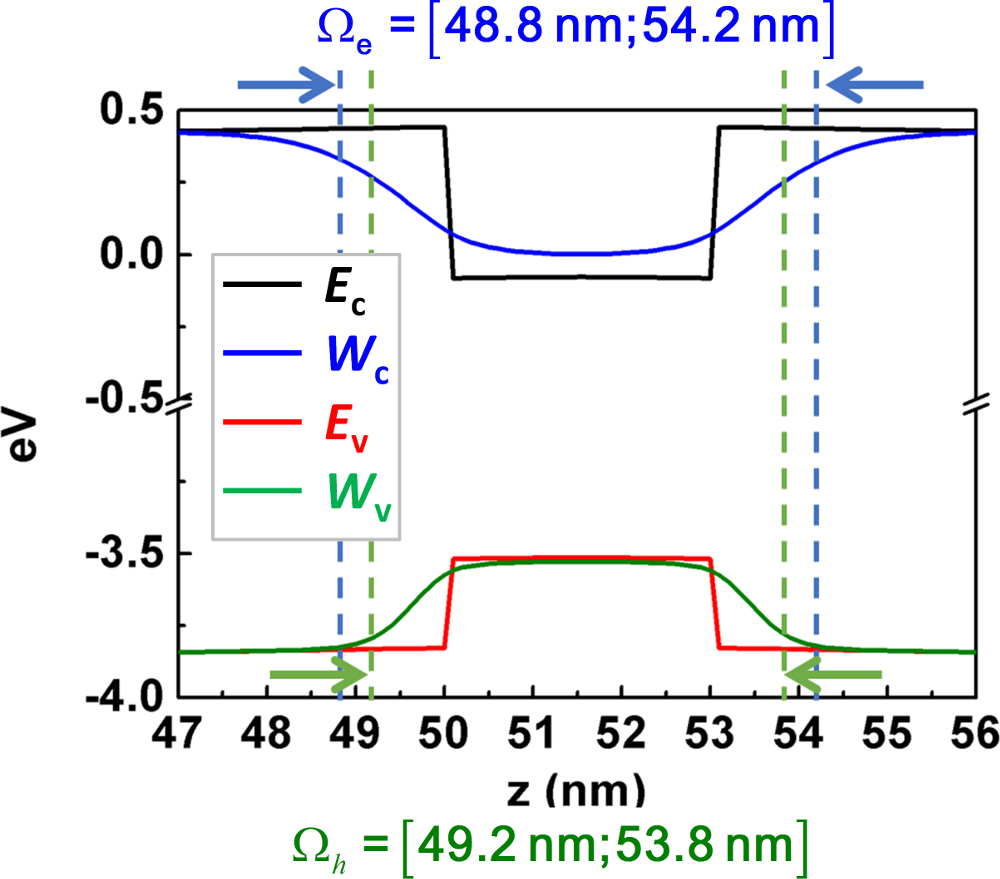}
\hskip 1mm
\includegraphics[width=0.23\textwidth]{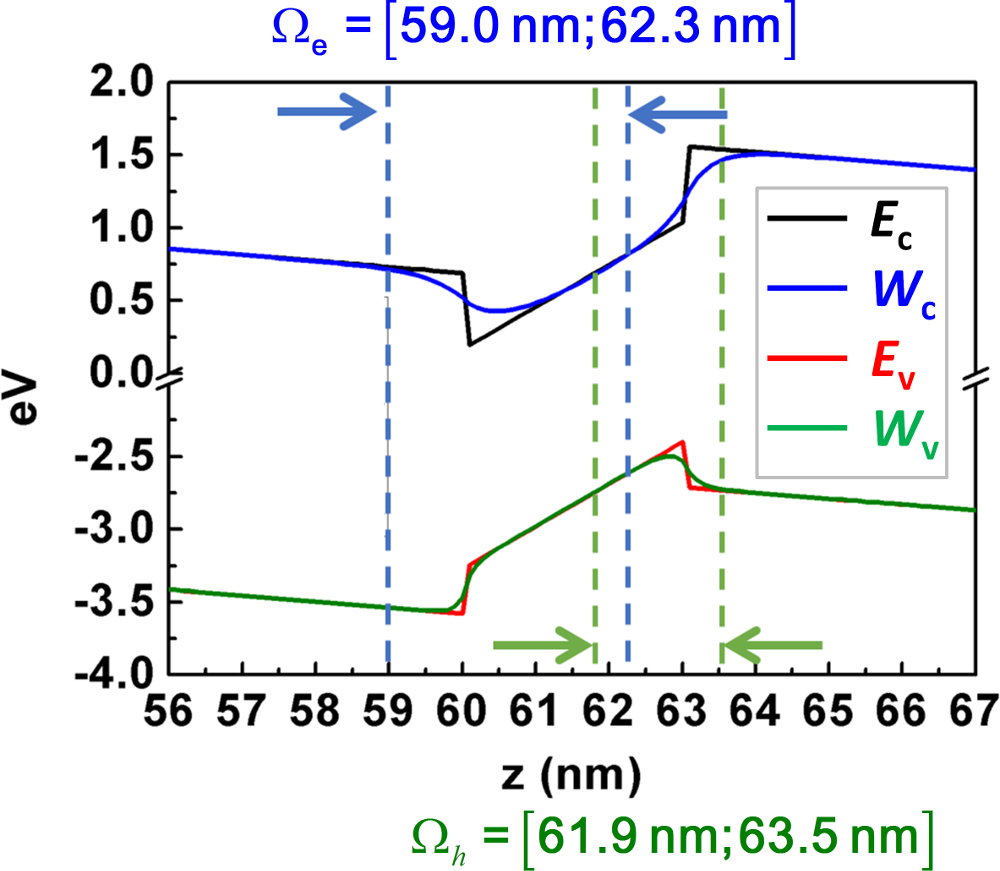}
\caption{Band structures for the $m$-plane single QW and the $c$-plane 3-QW. The localization potentials $W_c$ (blue) and $W_v$ (green) are superimposed over the band edges $E_c$ (black) and $E_v$ (red). The integration regions $\Omega_e$ and $\Omega_h$ are indicated on each frame.
\label{fig:MQW_bands}}
\end{figure}

The LLs and the corresponding localization potentials are computed from the conduction and valence bands for both structures without external applied bias. Figure~\ref{fig:MQW_bands} displays the corresponding band structures, together with the effective potentials $W_c$ and $W_v$. Despite the sharp boundaries of the conduction and valence bands at the well-barrier interface, one can see that the variations of the potentials $W_c$ and $W_v$ extend much further than the width of the well. The fundamental energies for electrons and holes in the well are therefore computed from Eq.~(\ref{eq:eigen_energy}) on a larger integration region than the well itself. The integration boundaries (different for electron and hole) are represented by the dashed lines in Fig.~\ref{fig:MQW_bands}. The boundaries of the integration region outside from the~QW are first set to the inflection points of the local effective potential~$W$. Then the integration boundaries are extended from these points by the characteristic decay length of a wave function of energy $E$ in a barrier of height~$W$, i.e., $\hbar/\sqrt{2m(W-E)}$. The lighter effective mass of electrons therefore leads to a larger integration domain than the one of the holes which have a heavier effective mass. Eigen-energies as well as the overlap between electron and hole fundamental states $\psi_{e}$ and $\psi_{h}$ are computed on these integration regions, $\psi_e$ and $\psi_h$ being approximated by $u_e$ and $u_h$ normalized on the same region (cf. Eq.~(\ref{eq:u_groundstate2})).

\begin{figure}
\includegraphics[width=0.45\textwidth]{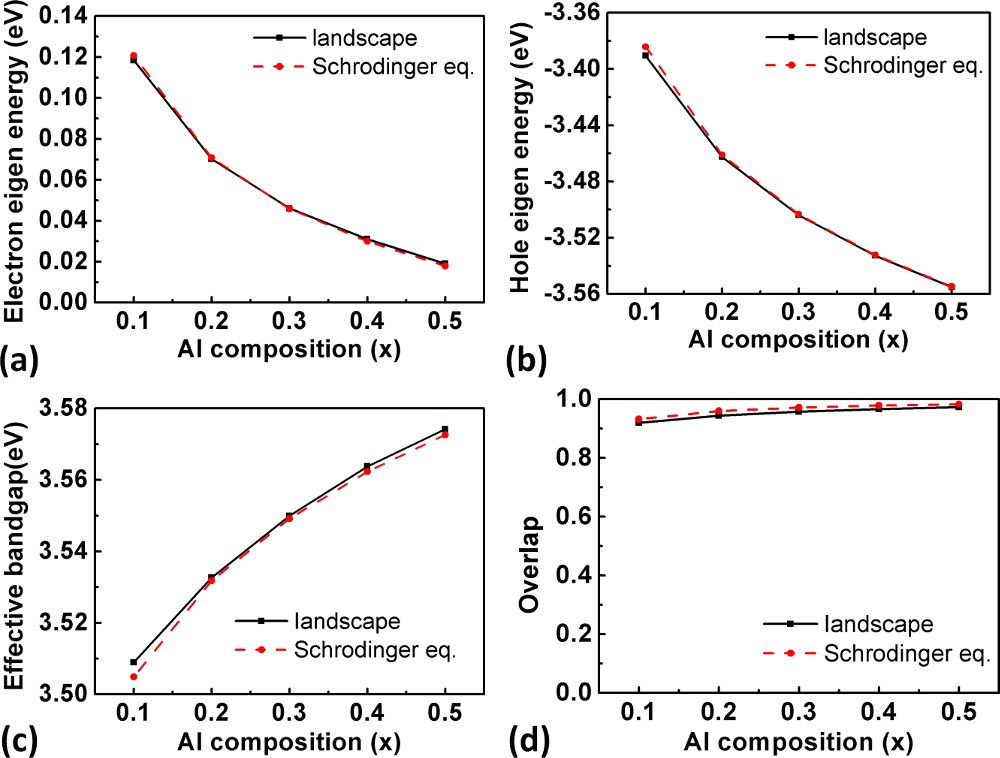}
\caption{Simulations results for the $m$-plane SQW (left structure in Fig.~\ref{fig:MQW_structures}). The energies computed by solving directly the Schr\"odinger equation are compared to the energies computed using our Poisson-LL model. Top frames display comparisons for electrons (left) and holes (right), respectively. Bottom left frame displays the energy difference, i.e. the smallest energy for a radiative transition. Bottom right frame shows the overlap integral between fundamental electron and hole states, computed from the Schr\"odinger equations (dashed line) and from the estimate using the LLs (plain line).
\label{fig:MQW_mplane_energies}}
\end{figure}

\begin{figure}
\includegraphics[width=0.45\textwidth]{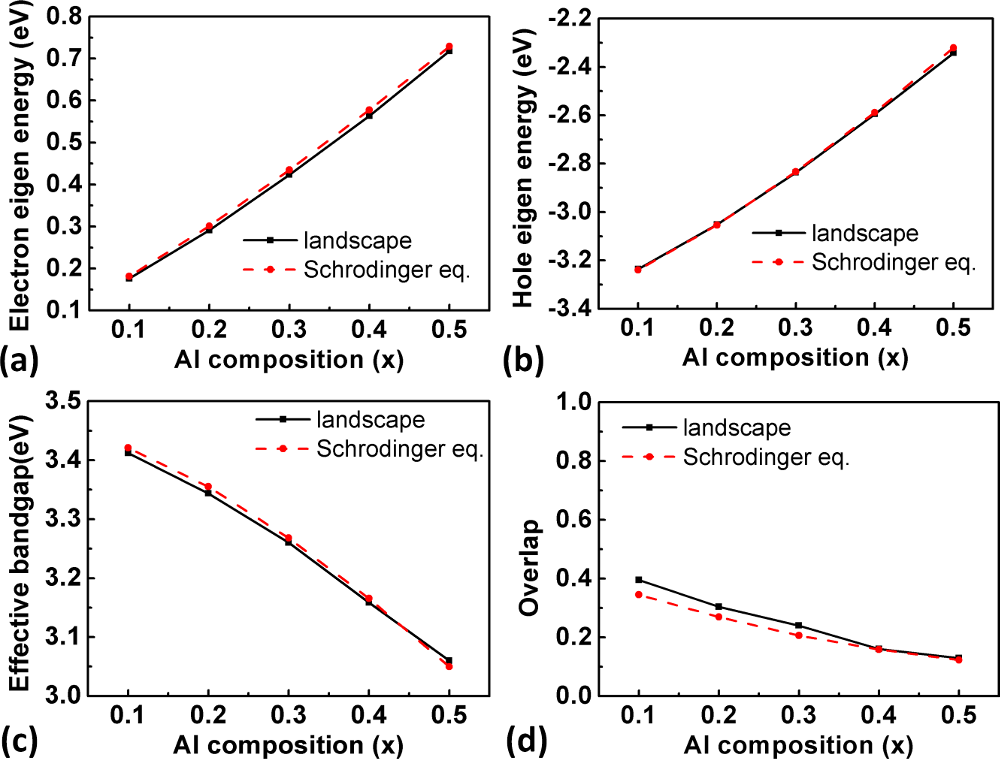}
\caption{Simulations results for the $c$-plane 3-QW (right structure in Fig.~\ref{fig:MQW_structures}). The energies computed by solving directly the Schr\"odinger equation are compared to the energies computed using our Poisson-LL model. Top frames display comparisons for electrons (left) and holes (right), respectively. Bottom left frame displays the energy difference, i.e. the smallest energy for a radiative transition. Bottom right frame shows the overlap integral between fundamental electron and hole states of the central well, computed from the Schr\"odinger equations (dashed line) and from the estimate using the LLs (plain line).
\label{fig:MQW_cplane_energies}}
\end{figure}

Figures~\ref{fig:MQW_mplane_energies} and \ref{fig:MQW_cplane_energies} display the comparisons between the fundamental energies computed by solving the Schr\"odinger equation directly, and the energies computed using Eq.~(\ref{eq:eigen_energy}), for $m$-plane and $c$-plane cases with various values of the aluminum content. In both figures, the top frames exhibit a remarkable agreement between the two different computations. The bottom left frames show the differences between the electron and hole energy, in other terms the energy of the smallest radiative transition. The bottom right frames display the value of the overlap integral between electron and hole fundamental states (in the case of the 3-QW structure, the states are taken in the central quantum well) (dashed line), compared with the estimates obtained using Eq.~(\ref{eq:u_groundstate1}) for the wave functions. The overlap region used for the calculation are the intersections of the regions $\Omega_e$ and $\Omega_h$ displayed in Fig.~\ref{fig:MQW_bands}. Although one observes here a slight deviation by a few percents of the approximated value using the landscapes, the agreement, both in absolute value and in trend, remains very good in the two structures, for all values of the Al content.

\subsection{Density of states, overlap and absorption computation}
\label{sub:SL_DOS}

\subsubsection{Periodic superlattice}

AlGaN/GaN superlattice (SL) structures are widely used in commercial LEDs to prevent the electron current leakage and improve lateral current spreading.\cite{Han2009, Song2013} However, modeling SL structures still remains a challenge for classical Poisson-DD solvers. In the classical picture, the resistance experienced by carriers is determined by the barrier height. In SLs, this leads to an overestimation of its value. Schr\"odinger-based solvers can model SLs, accounting for wave function coupling and tunneling effects. However, this approach cannot be applied in the case of multidimensional devices to study current crowding effects of disordered systems, due to the high demand of computation time. We show here that the LL~theory allows us to overcome this constraint. 

\begin{figure}
\includegraphics[width=0.38\textwidth]{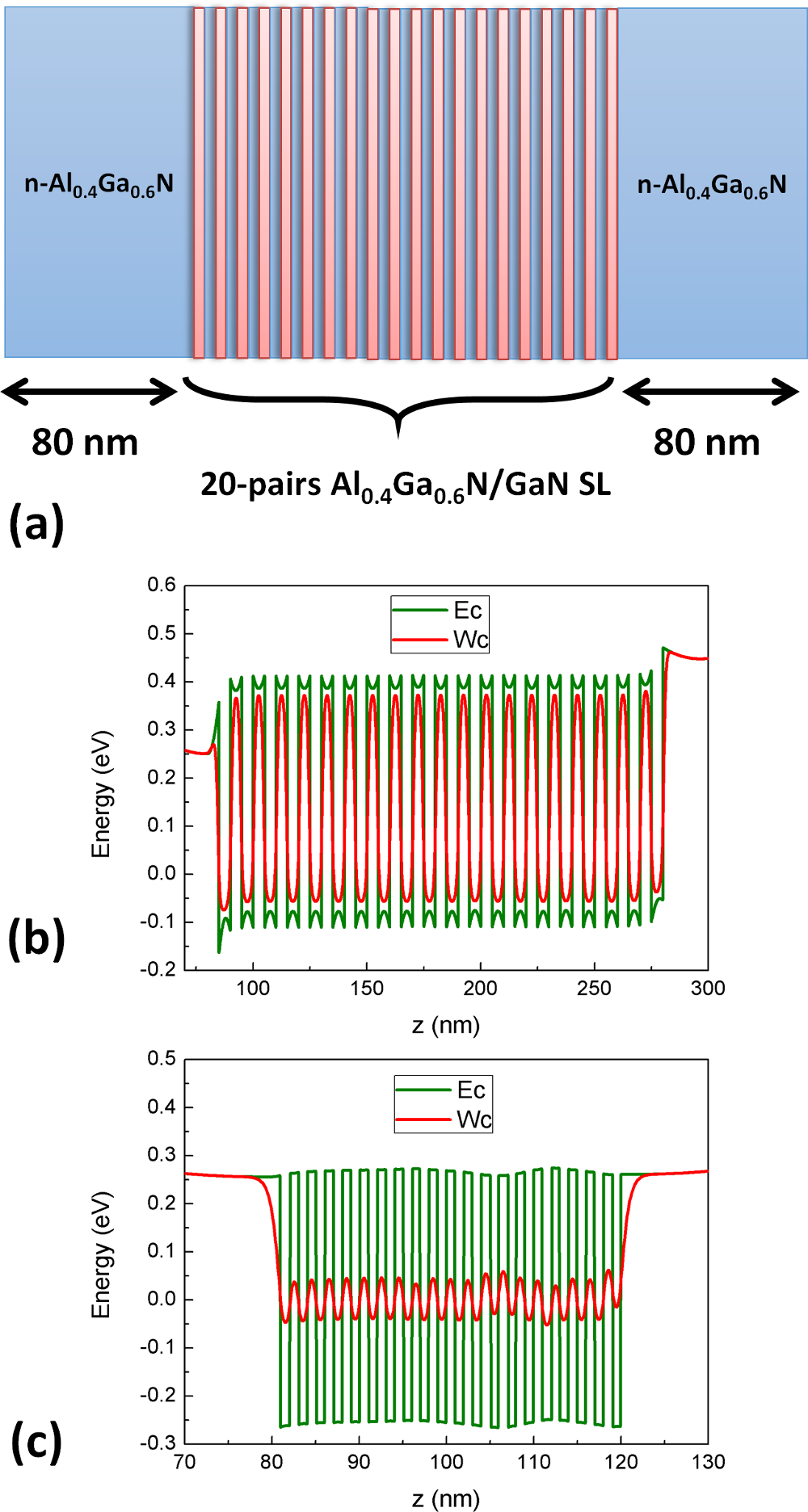}
\caption{(a) Schematic structure of 20-period Al$_{0.4}$Ga$_{0.6}$N/GaN SLs. (b),~(c) The conduction band potential ($E_c$) and effective quantum confining potential ($W_c$) for 5~nm/5~nm and 1~nm/1~nm SLs.
\label{fig:SL_structure}}
\end{figure}

We model a 20-pairs $n$-type $m$-plane Al$_{0.4}$Ga$_{0.6}$N/GaN SL structure with flat band conditions at both ends, as shown in Fig.~\ref{fig:SL_structure}(a), to analyze the electron transport behavior. Two structures with different periodicities, 5~nm/5~nm and 1~nm/1~nm QW/barrier thickness, are simulated. A nonpolar $m$-plane orientation is considered at first to study the intrinsic transport properties of SLs as described by the landscape model. In the thicker SL (5~nm/5~nm), quantum effects are weaker and $W_c$ deviates only slightly from $E_c$, as shown in Fig.~\ref{fig:SL_structure}(b). In contrast, in the thinner SL (1~nm/1~nm) $W_c$ is significantly different from $E_c$, as shown in Fig.~\ref{fig:SL_structure}(c). This difference is a manifestation of the coupling between wells when the barriers are thin, and expresses how the effective potential is able to translates quantum tunneling into a shift of the effective conduction band edges.

\begin{figure}
\includegraphics[width=0.33\textwidth]{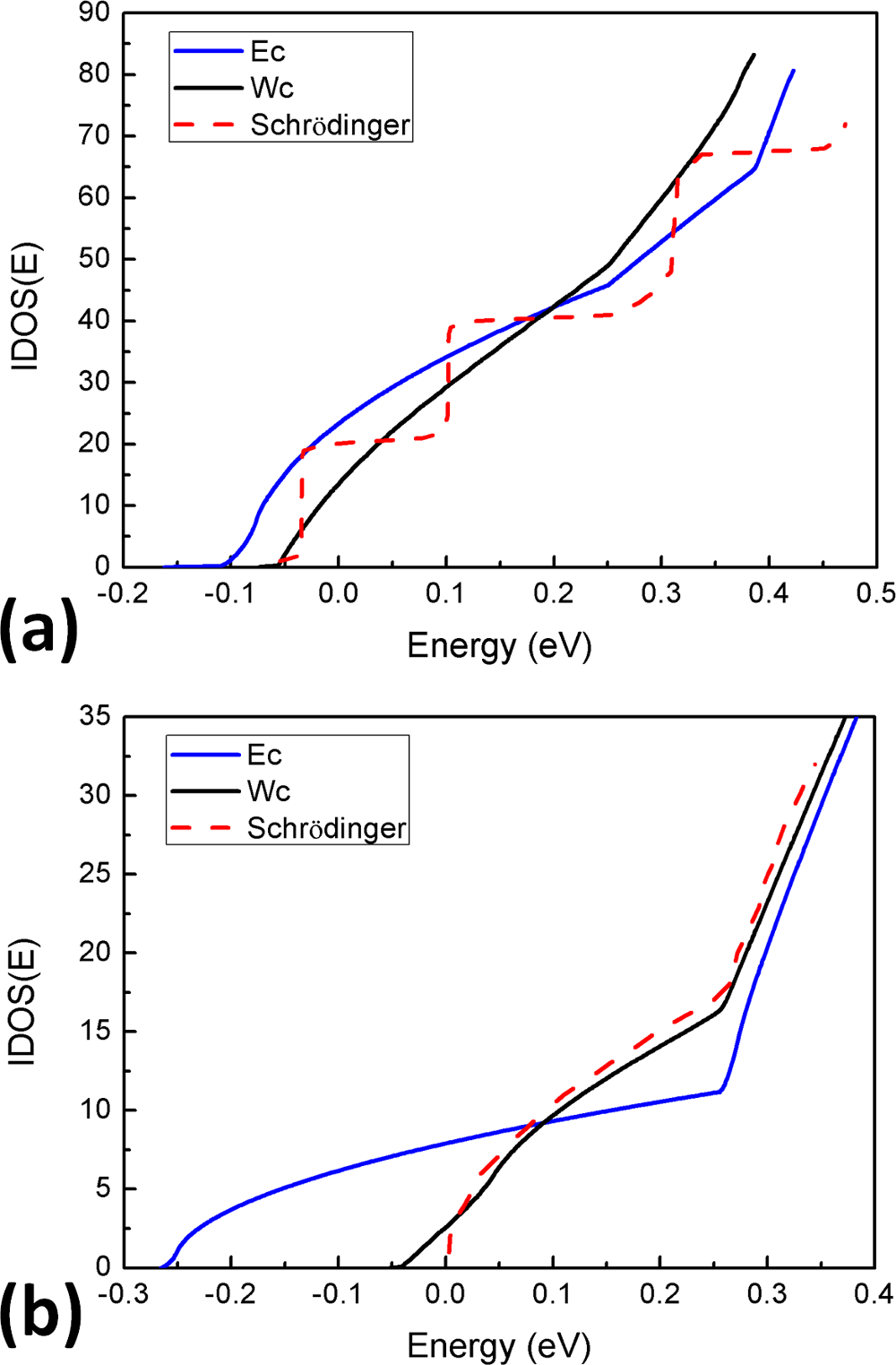}
\caption{(a),~(b) The counted states for 5~nm/5~nm and 1~nm/1~nm SLs.
\label{fig:SL_states}}
\end{figure}

To evaluate the effective barrier seen by carriers in a complicated structure such as SLs, Weyl's asymptotic formula is used~\cite{Arnold2016} to obtain the 1D LIDOS [see Eq.~(\ref{eq:N_E_1D})], which is then integrated over the entire system:
\begin{equation}
\rm IDOS(E) = \frac{2}{\pi} \int_0^L \sqrt{\frac{2 m_e^*(E-W_c(z))}{\hbar^2}}~dz\,
\end{equation}
where IDOS$(E)$ is the integrated density of available states. As seen in Section~\ref{subsec:effective_potential}, $W_c(z)=1/u_e(z)$ can be understood as the effective conduction band edge.

For large SL barrier and QW thicknesses, quantum effects are weak and classical. Therefore original potential- and landscape-based models give very similar estimates of the real IDOS (blue and black lines), as displayed in Fig. 11(a). One can observe that the take-off energy is better approximated by the landscape model (continuous black line) when compared to the computation from Schr\"odinger equation (red dashed line). In this situation of very weak coupling between wells, the similarity of all wells (no disorder here) induces a strong energy degeneracy which appears in the real IDOS. This IDOS increases through large steps of height 20 (the number of wells) which underlines the discrete nature of the spectrum, while the landscape-based (or $W_c$-based) approximation shows a continuous line following these steps.

As the SL thicknesses of QWs and barriers becomes smaller, wells become coupled, the degeneracy is lifted, creating subbands in the entire structure, and the corresponding IDOS (red dashed line) proceeds by smaller steps of height~1 which are almost invisible at the scale of the figure. The landscape-based IDOS ($W_c$-based Weyl’s law) shows here an excellent agreement with the exact calculation from the Schr\"odinger equation [Fig.~\ref{fig:SL_states}(b)], showing that it takes well into account the coupling between wells, in other words, the tunneling phenomenon. The approximation based on the original potential $E_c$, however, falls very far off the true IDOS, especially for lower energy states. Going into finer detail, one can also observe that in this last case, the $W$~approximation of the IDOS becomes positive about 50~meV earlier than the actual IDOS of the Schr\"odinger equation. This slight discrepancy comes from the fact that, by definition, the $W$~approximation of the IDOS is continuous while the actual IDOS is step-wise. Therefore, to reach a value close to~1 at an energy corresponding to the fundamental energy of the system, the approximate IDOS has to take off and be positive at a smaller energy.

\begin{figure}
\includegraphics[width=0.45\textwidth]{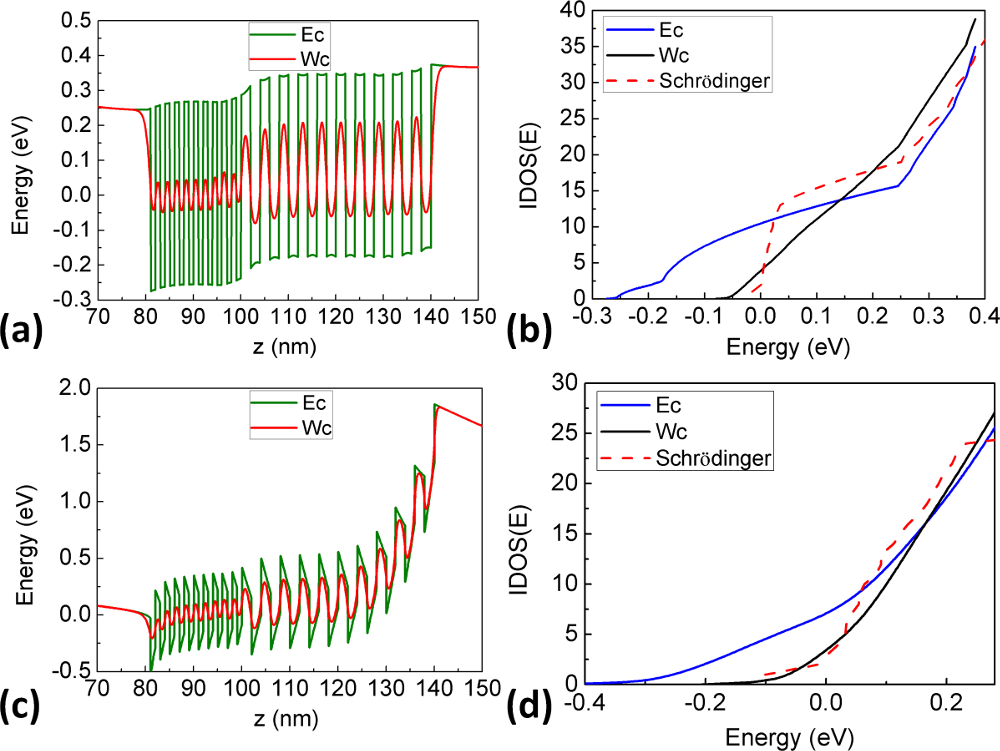}
\caption{(a) and (c) are the conduction band potential ($E_c$) and effective quantum confining potential ($W_c$) for $m$-plane (top) and $c$-plane (bottom) double period SLs. (b),~(d) The corresponding integrated density of states, IDOS.
\label{fig:SL_bands}}
\end{figure}

To test further the LL~model, we analyze more complicated structures such as 20-pairs $n$-type SL with double periods composed of 10-pairs 1~nm/1~nm and 10-pairs 2~nm/2~nm SLs, respectively, including $m$-plane and $c$-plane cases. Figures~\ref{fig:SL_bands}(a) and \ref{fig:SL_bands}(c) show the conduction band edge~$E_c$ and $W_c$ for $m$-plane and $c$-plane conditions. Here again, the difference between $E_c$ and $W_c$ increases as the period become smaller, exhibiting stronger tunneling effects. The IDOS displayed in Figs~\ref{fig:SL_bands}(b) and \ref{fig:SL_bands}(d) shows the quality of the approximation provided by $W_c$, regardless of the structure complexity. We found the $c$-plane case exhibits a better agreement with the Schr\"odinger model due to distinct potential energy distributions. As randomness and inter-coupling increase in the system, the prediction of $W_c$ becomes closer to the solution of the Schr\"odinger equation. This last result was in fact already partly demonstrated in~Ref.~\citenum{Arnold2016} for very random systems.

\subsubsection{Disordered superlattice}
\label{subsub:DSL}

To illustrate the efficiency of the landscape model in the case of disordered systems, we compute the band structure and the density of states in nitride $m$-plane and $c$-plane disordered SL. Two types of SL are investigated: the first is composed of 20 pairs of well/barrier layers, while the second is composed of 200 pairs. In both cases, wells and barriers have 1-nm thickness. The barrier material is  Al$_{0.4}$Ga$_{0.6}$N and the well material is Al$_x$Ga$_{1-x}$N, where $x$ is randomly and independently determined in each well, using a uniform law between 0 and 0.4.

\begin{figure}
\includegraphics[width=0.48\textwidth]{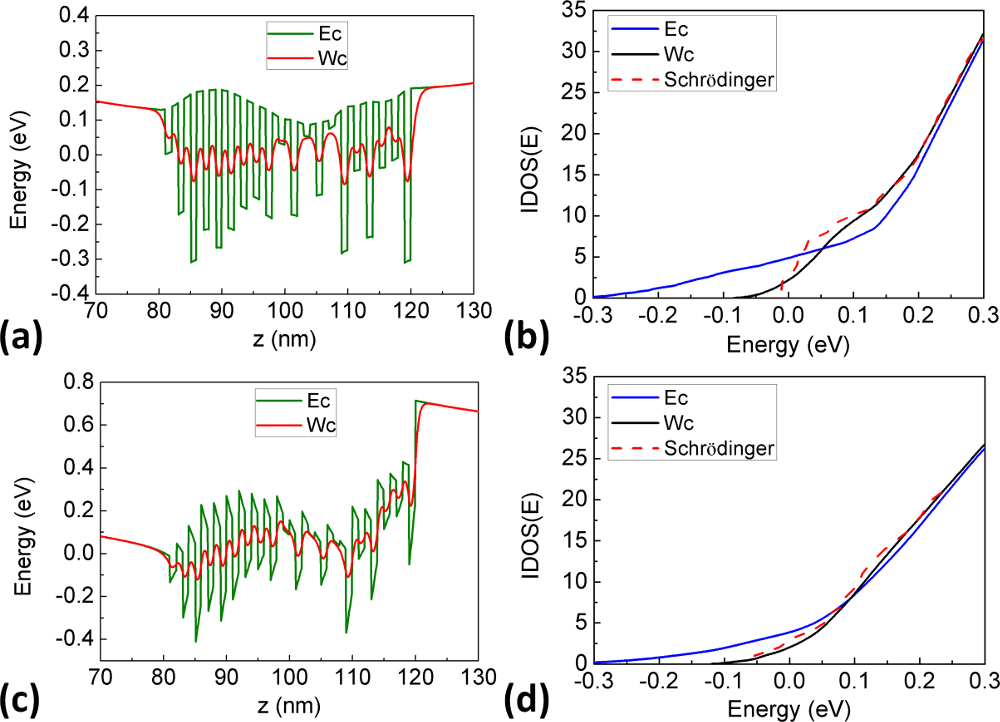}
\caption{(a),~(c) The conduction band potential ($E_c$) and effective quantum confining potential ($W_c$) for $m$-plane (top) and $c$-plane (bottom) 20-pairs disordered SL. (b),~(d) The corresponding IDOS.}
\label{fig:DSL_20}
\end{figure}

\begin{figure}
\includegraphics[width=0.48\textwidth]{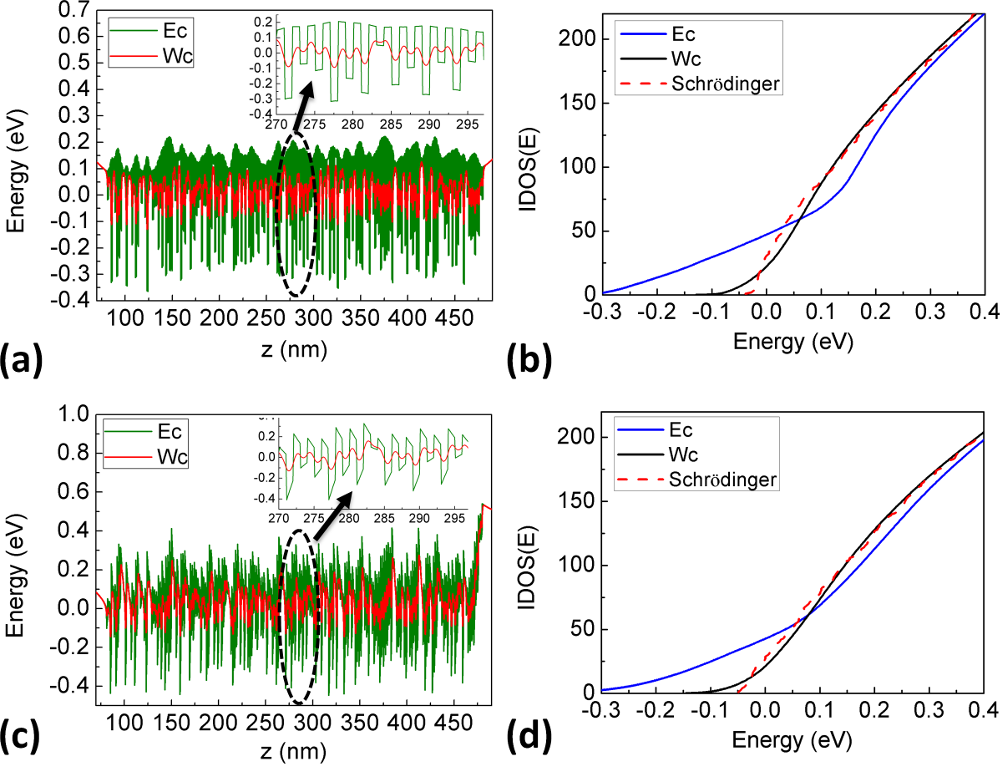}
\caption{(a),~(c) The conduction band potential ($E_c$) and effective quantum confining potential ($W_c$) for $m$-plane (top) and $c$-plane (bottom) 200-pairs disordered SL. (b),~(d) The corresponding IDOS.}
\label{fig:DSL_200}
\end{figure}

For each type of SL, the band structure and the IDOS are computed (Figs~\ref{fig:DSL_20} and \ref{fig:DSL_200}). In both cases, although the edge of the conduction band $E_c$ now exhibits large fluctuations across the structure due to the compositional disorder, the quantum coupling between wells translates into a much smoother effective potential~$W_c$. The value of $W_c$ is significantly larger than $E_c$ and can be interpreted as a local fundamental energy from the expression of the local density of states in Eq.~(\ref{eq:LDOS_1D}). Looking at the IDOS (right column of Figs~\ref{fig:DSL_20} and \ref{fig:DSL_200}), we observe a very good agreement between the actual counting functions and its approximation obtained using~$W_c$. One has to note the same slight discrepancy of the take-off energy between both functions that was observed in the periodic case. This discrepancy, which comes from the different natures of the two functions (continuous versus stepwise), appears to be of the same order of magnitude independently on the number of wells in the structure.

\begin{figure}
\includegraphics[width=0.3\textwidth]{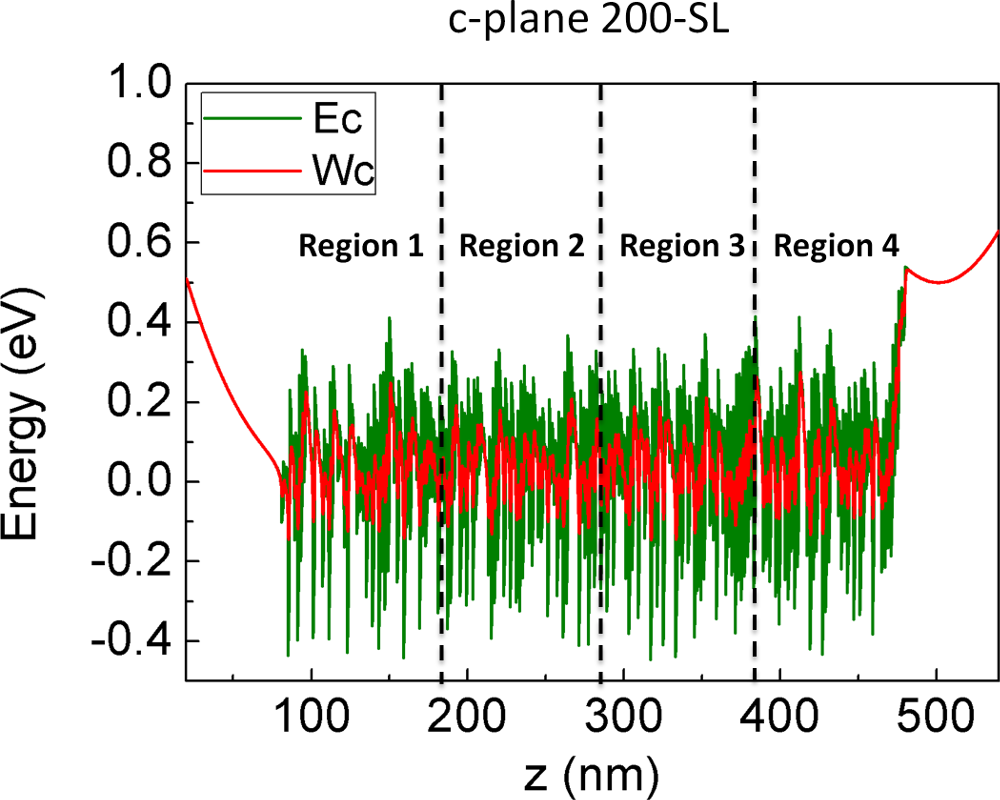}
\vskip 2mm
\includegraphics[width=0.48\textwidth]{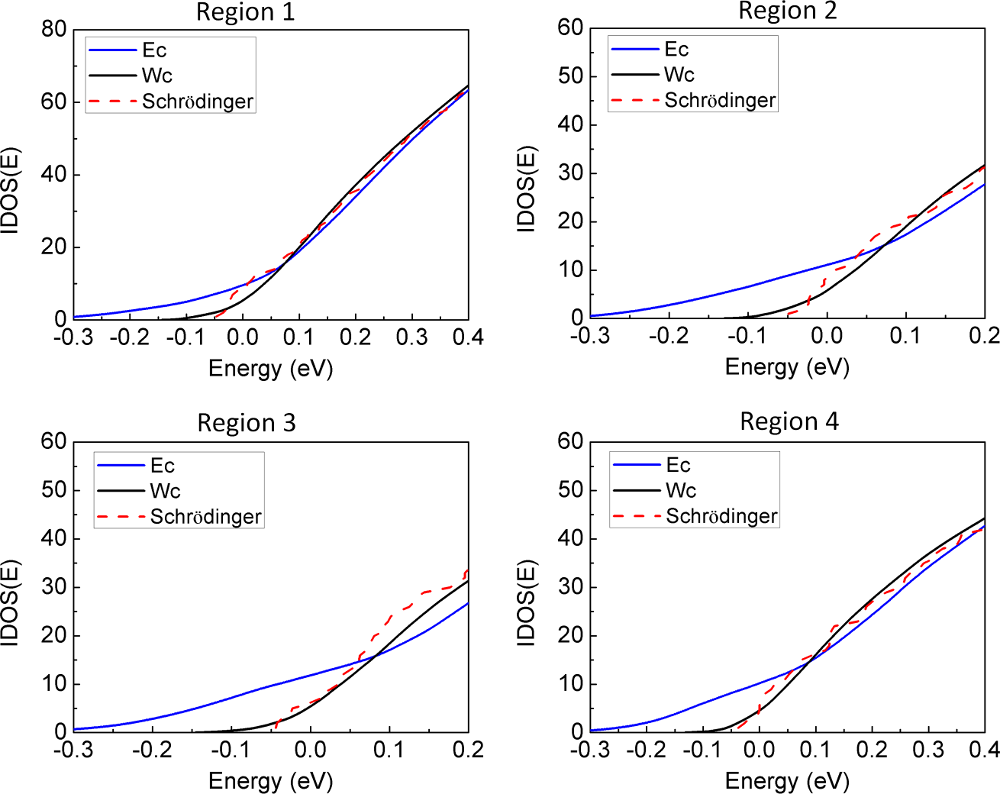}
\caption{Comparison of the LIDOS of a $c$-plane 200-pairs disordered SL structure computed using on one hand a coupled Poisson-Schr\"odinger approach, and on the other hand our coupled Poisson-LL model. The structure is divided into four regions (top frame). Comparisons of the IDOS of each region computed using both methods are displayed in the four bottom frames.}
\label{fig:SL200_4regions_LIDOS}
\end{figure}

We also use the $c$-plane 200-pairs disordered SL structure (bottom of Fig.~\ref{fig:DSL_200}) to test the quality of the local integrated density of states approximation. To this end, the SL structure is divided into four disjoint regions of same length. The local integrated density of states (LIDOS) of each region is computed in two different ways: first, by directly solving the Schr\"odinger equation in the entire system, each quantum state is then assigned a unique region among the four already defined, which is the region where the state reaches its maximum amplitude. Secondly, by integrating the LDOS of Eq.~(\ref{eq:LDOS_1D}) on each regio, comparisons of the two methods for all four regions are presented in the four bottom frames of Fig.~\ref{fig:SL200_4regions_LIDOS}. Here also one can observe the accuracy of the LIDOS computed using the landscape approach.

\begin{figure}
\includegraphics[width=0.48\textwidth]{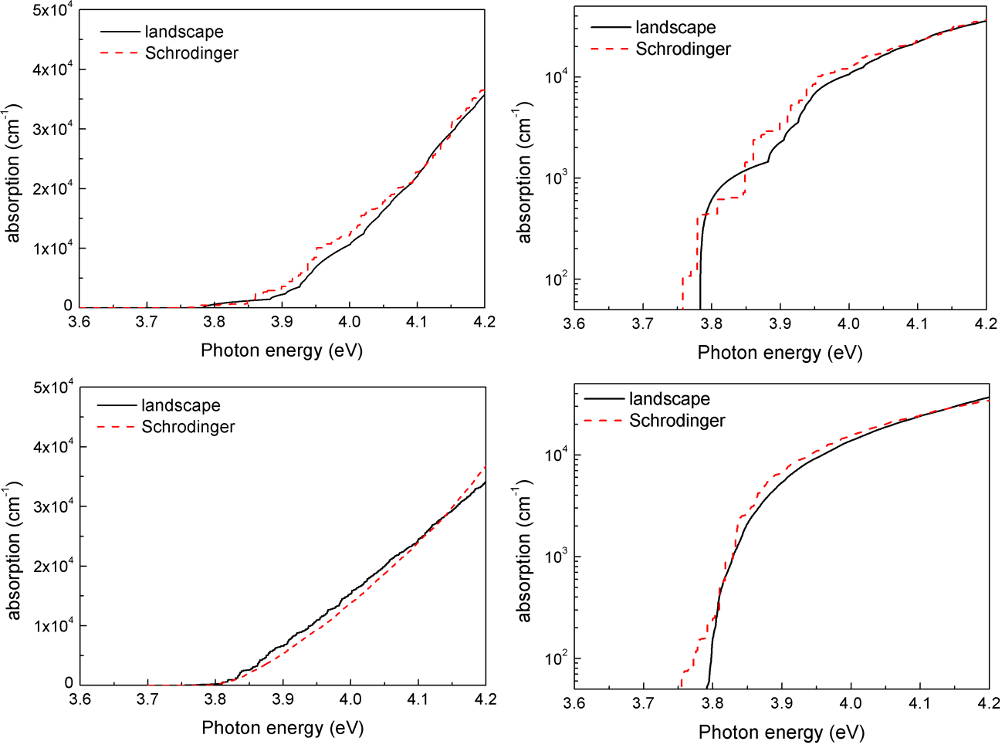}
\caption{(Top) Light absorption spectra (in linear and log scales) for the $c$-plane 20-pairs disordered SL of Fig.~\ref{fig:DSL_20}. The absorption spectrum computed using the Schr\"odinger equation is displayed with a red dashed line while the spectrum computed with the landscape approach is displayed with a black continuous line. (Bottom) Light absorption spectrum (in linear and log scales) for the $c$-plane 200-pairs disordered SL of Fig.~\ref{fig:DSL_200}.}
\label{fig:absorptions}
\end{figure}

We finally test the accuracy of the overlap and joint density of states estimates. To this end, the absorption spectrum between electrons and heavy holes (of respective masses 0.20 and 1.87) of both disordered SL (20 and 200 wells) are computing summing over all localization subregions the formula of Eq.~(\ref{eq:absorption_disorderedQW_landscape}), and compared to exact computations using the quantum states computed by Schr\"odinger equation. Comparisons between the two types of calculations are displayed in Fig.~\ref{fig:absorptions}, both in linear and log scale. One can see that the absorption curves obtained using the LL (plain black lines) are very close to the ones computed using Schr\"odinger equation (red dashed lines), over almost three decades of absorption rate, confirming the quality of the landscape approach. Our results show that $W$ appears as a very likely candidate for inserting into a model of carrier transport able to account for quantum effects in complicated systems.

\subsection{Carrier distribution}
\label{sub:carrier_comparison}

To study and compare in detail the carrier distributions predicted by the different models, a 1D~SQW is simulated. The structure is composed of a 3-nm active GaN layer, enclosed between two barriers of Al$_x$Ga$_{1-x}$N, where $x=0.2$ will be the reference case. The band offsets between GaN/AlGaN conduction bands are assumed to be 63$\%$ of the band-gap discontinuity. The detailed band structure parameters for GaN and AlN are provided in Table~\ref{tab:parameter} (see Appendix). All parameters of Al$_{x}$Ga$_{1-x}$N alloys are obtained by interpolation, where the bandgap alloy bowing parameter is assumed to be 0.8~eV.

\begin{figure}[h!]
\includegraphics[width=0.4\textwidth]{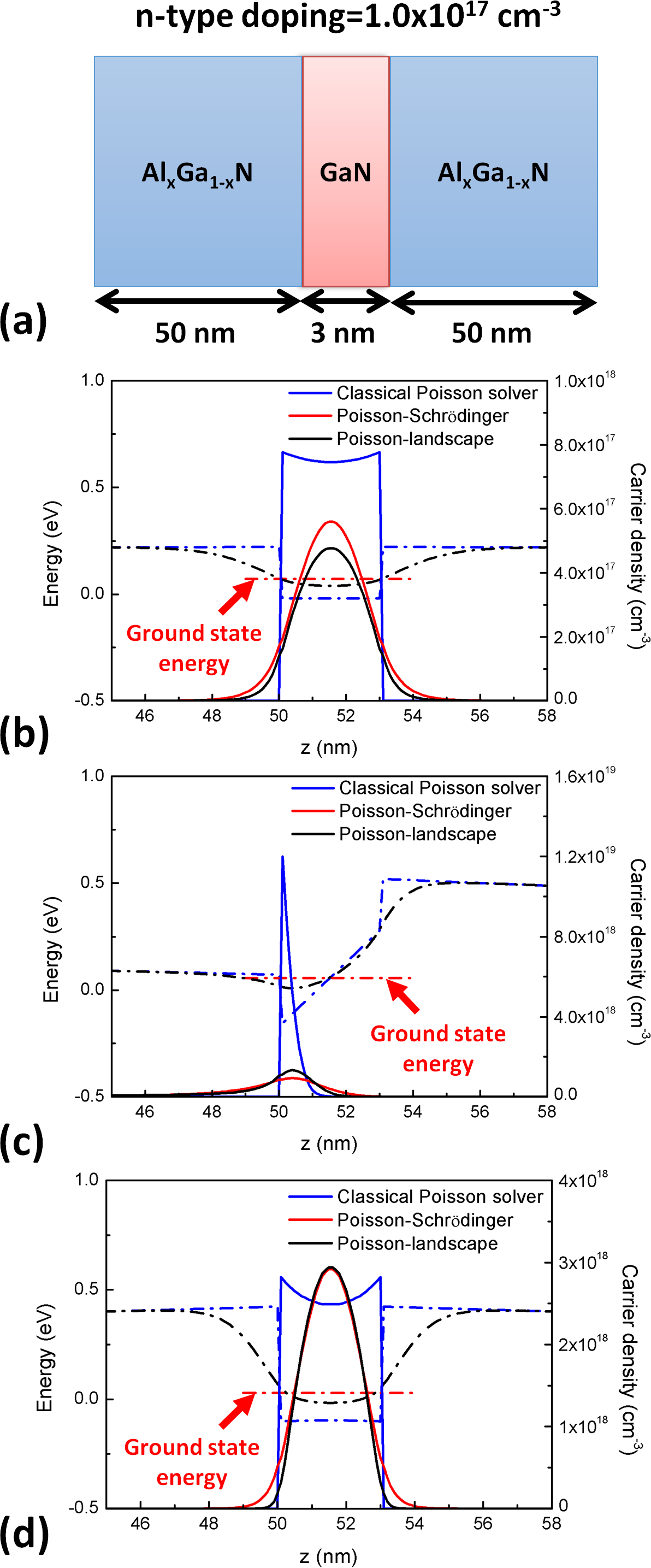}
\caption{(a) Schematic structure of the single Al$_x$Ga$_{1-x}$N/GaN/Al$_x$Ga$_{1-x}$N QW, where $x$ is variable. (b) Potential energy and carrier distribution for $m$-plane and $x=0.2$. (c) Potential energy and carrier distribution for $c$-plane and $x=0.2$. (d) Potential energy and carrier distribution for $m$-plane and $x=0.4$. The blue, black, and red lines are solved by the classical Poisson equation, Poisson-LL model, and Poisson-Schr\"odinger equation, respectively. The Fermi level is located at zero energy as the reference.\label{fig:1D_SQW}}
\end{figure}

The $m$-plane and $c$-plane cases (without and with the polarization charge induced at the interface) are both discussed. The detailed dimension and material doping level are shown in Fig.~\ref{fig:1D_SQW}(a), where the electron effective mass of Al$_{0.2}$Ga$_{0.8}$N and GaN is 0.214m$_0$ and 0.20m$_0$, respectively. The doping is assumed to be fully activated to ensure the same activation condition for comparison.

The carrier density distribution is solved through three different methods: (1) the classical Poisson model solving Poisson equation for the charge distribution of ionized donors across the QW structure; (2) self-consistent Poisson-LL approach; and (3) self-consistent Poisson-Schr\"odinger approach. Figure~\ref{fig:1D_SQW}(b) displays the computed carrier distribution and the potential energy for the $m$-plane case with a symmetric potential energy by the three methods. We can see that the carrier distribution obtained from the classical Poisson equation (blue line) is almost constant across the well and dropping sharply outside, not accounting for the quantum nature of the electron. The Poisson-Schr\"odinger solver provides a very different outcome: the carrier distribution is more confined in the center of the well [red line of Fig.~\ref{fig:1D_SQW}(b)] and extends smoothly outside of it, revealing the wave-function shape of the carrier. Turning to the landscape model, the effective quantum potential~$W_c$ (black line) exhibits a smoother behavior than the original conduction band (blue line). The carrier density computed after replacing the original conduction band edge~$E_c$ with $W_c$ appears very similar to that computed from the Schr\"odinger equation.

In Fig.~\ref{fig:1D_SQW}(c), we evaluate the $c$-plane case with an asymmetric potential profile induced by the polarization charge. The carrier distribution computed by the classical Poisson equation is sharp and mostly located at the minimum of the conduction band. This does not match the result obtained from the Poisson-Schrodinger solver. This excessively large carrier density located at the interface might be the reason why some numerical studies\cite{Kuo2010} using Poisson and DD solvers to study GaN-based polar QWs adopt only $\sim$50\% of the theoretical polarization charge, in order to reproduce a carrier distribution consistent with the Schr\"odinger equation. Here again, the Poisson-LL model results in a smoother carrier distribution, as shown in Fig.~\ref{fig:1D_SQW}(c), much closer to the Schr\"odinger solution.

\begin{figure}
\includegraphics[width=0.35\textwidth]{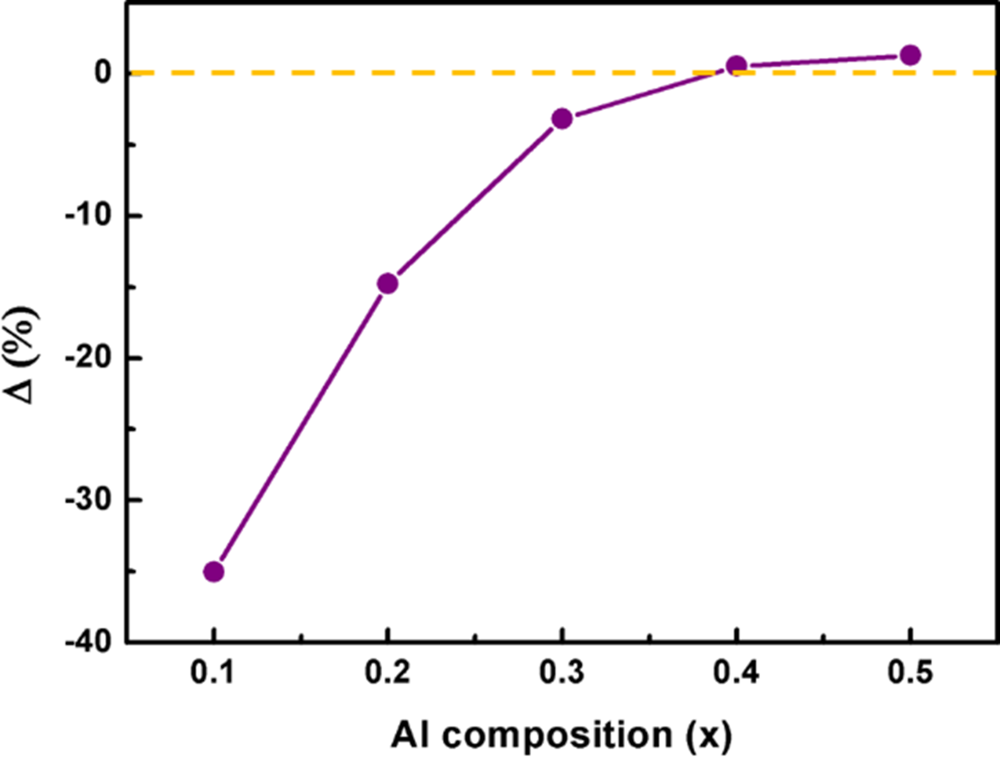}
\caption{Relative difference between the predictions of the peak carrier distribution by the Poisson-LL and Poisson-Sch\"odinger models for various Al compositions of the Al$_x$Ga$_{1-x}$N/GaN $m$-plane SQW structure.
\label{fig:Delta}}
\end{figure}

In the case of a deeper $m$-plane QW with $x=0.4$, carrier distributions computed with the Poisson-LL and Poisson-Schr\"odinger models are also in good agreement, as displayed in Fig.~\ref{fig:1D_SQW}(d). To quantify this agreement, we simulate a series of Al composition from 10\% to 50\% ($0.1 \le x \le 0.5$). For each simulation, we compute the dimensionless quantity $\Delta$ defined as the relative difference of peak carrier densities between the Poisson-landscape and Poisson-Schr\"odinger models. As displayed in Fig.~\ref{fig:Delta}, $\Delta$ becomes smaller when the carriers are well localized within a deeper potential (larger Al composition) and almost vanishes above 40\% Al composition.

We can conclude from this that the LL~model matches very well the solution of the Schr\"odinger equation when the system is strongly localized. Even in systems with a lower degree of localization (here, small $x$) the prediction of the Poisson-LL model gives an acceptable agreement with the exact Poisson-Schr\"odinger solution and provides an overall description of the carrier distribution much more accurate than the result of a classical Poisson solver. The discrepancy observed between the carrier distributions computed in Poisson-LL and Poisson-Schr\"odinger models can be attributed to the fact that the local DOS used in Poisson-LL does not fully describe the exact shape of the wave function. If required, an even more accurate approximation might be achieved by using the effective confining potential to compute a correction to this carrier distribution without any adjustable parameter.\cite{Ancona1987, Ancona1989}

\section{Conclusion}

In this work, we applied the LL~theory,\cite{Filoche2012, Arnold2016} until now a purely mathematical framework, to build a new model of quantum and disordered semiconductors devices. In the LL~theory, the Schr\"odinger equation is replaced by an associated Dirichlet equation whose solution is called the localization landscape, $u$. The reciprocal of this landscape, $W=1/u$, acts as an effective quantum classical confining potential which governs localization of quantum states. Eigenfunction profiles and geometry of localization subregions can be retrieved from a direct analysis of this effective potential. Inserting~$W$ into Weyl's law also provides a very good approximation of the density of states, the carrier concentrations, and the spatial distributions of charges, especially in the case of strong localization. Besides, as seen here in 1D cases, and shown to play a decisive role in the 3D modeling of LEDs (see LL3, Ref.~\citenum{Li2017}), the LL~theory simulates to an excellent approximation two major effects of quantum mechanics, namely, the reduction of barrier heights (tunneling effect) and the raising of energy ground states (quantum confinement effect).

We have presented here how to compute localized states, energies, density of states using the landscapes, and how to couple them to Poisson and DD equations to model carrier transport in SL structures and carrier localization in nitride-based systems. In principle, this method is not restricted to modeling nitride-based devices (as the examples presented here or in companion papers dealing with the simulation of the absorption edge~(LL2, Ref.~\citenum{Piccardo2017}) and LEDs~(LL3, Ref.~\citenum{Li2017}) based on InGaN alloy materials), but can expand to other semiconductor materials and to any electronic or optoelectronic properties requiring the knowledge of electron and hole quantum states. Moreover, except for the transport and carrier distribution issues discussed in this paper, the emission and absorption in the disordered system can be properly modeled in terms of eigen-energy calculation (LL2, Ref.~\citenum{Piccardo2017}). Finally, as observed in real modeling exercises (LL3, Ref.~\citenum{Li2017}), the computation time using the landscape model is considerably reduced compared to a conventional Schr\"odinger solver, which makes this model ideal for simulating and designing quantum 3D~real-world devices.

\appendix*
\section{Band structures and polarization in AlGaN alloys}

Table~\ref{tab:parameter} displays the detailed band structure parameters for GaN and AlN, respectively. The parameters of all Al$_{x}$Ga$_{1-x}$N alloys are obtained by interpolation, the bandgap alloy bowing parameter being taken equal to 0.8~eV. Spontaneous and piezoelectric polarizations are computed using Eqs.~(\ref{eq:psp}) and (\ref{eq:ppz}), while parameters of polarization values and piezoelectric coefficients are provided in Tables~\ref{tab:parameter-pol} and \ref{tab:parameter-piezo}, respectively.
\begin{equation}
\label{eq:psp}
P^{sp} = a~x+b~(1-x)+c~x(1-x)
\end{equation}
\begin{equation}
\label{eq:ppz}
P^{pz}=[\bf{e}]\cdot[\bf{\varepsilon}]= \left(
  \begin{array}{c}
    e_{15}~\varepsilon_{xz} \\
    e_{15}~\varepsilon_{yz} \\
    e_{31}~(\varepsilon_{xx}+\varepsilon_{yy})+e_{33}~\varepsilon_{zz}  
  \end{array}
  \right)
\end{equation}

\begin{table}[h!]
\begin{center}
\begin{tabular}{ccccccc}
\hline 
\hline
 & $E_g$  & $\varepsilon_r$ & $m_{e}^{\parallel}$ & $m_{e}^{\perp}$ & $m_{hh}$ & $m_{lh}$ \\
~units~ & (eV) & & ($m_0$) & ($m_0$) & ($m_0$) & ($m_0$)
\\ \hline
~GaN~  & 3.437 & 10.4  & 0.21 & 0.20 & 1.87 & 0.14 \\
~AlN~  & 6.0   & 10.31 & 0.32 & 0.30 & 2.68 & 0.26 \\
\hline
\hline 
\end{tabular} 
\end{center}
\caption{Band structure parameters for wurtzite GaN and AlN alloys: band gap, relative permittivity, effective masses.}
\label{tab:parameter}
\end{table}

\begin{table}[h!]
\begin{center}
\begin{tabular}{ccc}
\hline 
\hline
   a     &  b      &  c   \\ \hline
 -0.090 & -0.034  &  0.021 \\
\hline
\hline 
\end{tabular} 
\end{center}
\caption{Interpolation parameters for polarization in Al$_{x}$Ga$_{1-x}$N.}
\label{tab:parameter-pol}
\end{table}

\begin{table}[h!]
\begin{center}
\begin{tabular}{cccc}
\hline 
\hline
 & $e_{33}$  & $e_{31}$  & $e_{15}$  \\ 
 ~units~ & (C.cm$^{-2}$) & (C.cm$^{-2}$) & (C.cm$^{-2}$) \\ \hline
~GaN~  & 0.73 & -0.49 & -0.40 \\
~AlN~  & 1.55 & -0.58 & -0.48 \\
\hline
\hline 
\end{tabular} 
\end{center}
\caption{Piezoelectric coefficients for wurtzite GaN and AlN alloys.}
\label{tab:parameter-piezo}
\end{table}

\begin{acknowledgments}
This work was supported by the project CRIPRONI (ANR-14-CE05-0048-01/MOST-104-2923-E-002-004-MY3) of the French National Research Agency (ANR) and Taiwanese Ministry of Science and Technology (MOST). Svitlana Mayboroda is partially supported by the Alfred P. Sloan Fellowship, the NSF CAREER Award DMS-1056004, the NSF MRSEC Seed Grant, and the NSF INSPIRE Grant. Additional support for Marco Piccardo and Claude Weisbuch was provided by the DOE Solid State Lighting Program under Award \#DE-EE0007096.
\end{acknowledgments}

\bibliography{localization}

\begin{thebibliography}{35}%
\makeatletter
\providecommand \@ifxundefined [1]{%
 \@ifx{#1\undefined}
}%
\providecommand \@ifnum [1]{%
 \ifnum #1\expandafter \@firstoftwo
 \else \expandafter \@secondoftwo
 \fi
}%
\providecommand \@ifx [1]{%
 \ifx #1\expandafter \@firstoftwo
 \else \expandafter \@secondoftwo
 \fi
}%
\providecommand \natexlab [1]{#1}%
\providecommand \enquote  [1]{``#1''}%
\providecommand \bibnamefont  [1]{#1}%
\providecommand \bibfnamefont [1]{#1}%
\providecommand \citenamefont [1]{#1}%
\providecommand \href@noop [0]{\@secondoftwo}%
\providecommand \href [0]{\begingroup \@sanitize@url \@href}%
\providecommand \@href[1]{\@@startlink{#1}\@@href}%
\providecommand \@@href[1]{\endgroup#1\@@endlink}%
\providecommand \@sanitize@url [0]{\catcode `\\12\catcode `\$12\catcode
  `\&12\catcode `\#12\catcode `\^12\catcode `\_12\catcode `\%12\relax}%
\providecommand \@@startlink[1]{}%
\providecommand \@@endlink[0]{}%
\providecommand \url  [0]{\begingroup\@sanitize@url \@url }%
\providecommand \@url [1]{\endgroup\@href {#1}{\urlprefix }}%
\providecommand \urlprefix  [0]{URL }%
\providecommand \Eprint [0]{\href }%
\providecommand \doibase [0]{http://dx.doi.org/}%
\providecommand \selectlanguage [0]{\@gobble}%
\providecommand \bibinfo  [0]{\@secondoftwo}%
\providecommand \bibfield  [0]{\@secondoftwo}%
\providecommand \translation [1]{[#1]}%
\providecommand \BibitemOpen [0]{}%
\providecommand \bibitemStop [0]{}%
\providecommand \bibitemNoStop [0]{.\EOS\space}%
\providecommand \EOS [0]{\spacefactor3000\relax}%
\providecommand \BibitemShut  [1]{\csname bibitem#1\endcsname}%
\let\auto@bib@innerbib\@empty
\bibitem [{\citenamefont {Weisbuch}\ and\ \citenamefont
  {Vinter}(1991)}]{Weisbuch1991}%
  \BibitemOpen
  \bibfield  {author} {\bibinfo {author} {\bibfnamefont {C.}~\bibnamefont
  {Weisbuch}}\ and\ \bibinfo {author} {\bibfnamefont {B.}~\bibnamefont
  {Vinter}},\ }\href@noop {} {\emph {\bibinfo {title} {Quantum Semiconductor
  Structures}}}\ (\bibinfo  {publisher} {Academic Press},\ \bibinfo {address}
  {Boston},\ \bibinfo {year} {1991})\BibitemShut {NoStop}%
\bibitem [{\citenamefont {Alferov}(2001)}]{Alferov2001}%
  \BibitemOpen
  \bibfield  {author} {\bibinfo {author} {\bibfnamefont {Z.~I.}\ \bibnamefont
  {Alferov}},\ }\href {\doibase 10.1103/RevModPhys.73.767} {\bibfield
  {journal} {\bibinfo  {journal} {Rev. Mod. Phys.}\ }\textbf {\bibinfo {volume}
  {73}},\ \bibinfo {pages} {767} (\bibinfo {year} {2001})}\BibitemShut
  {NoStop}%
\bibitem [{\citenamefont {Kroemer}(2001)}]{Kroemer2001}%
  \BibitemOpen
  \bibfield  {author} {\bibinfo {author} {\bibfnamefont {H.}~\bibnamefont
  {Kroemer}},\ }\href {\doibase 10.1103/RevModPhys.73.783} {\bibfield
  {journal} {\bibinfo  {journal} {Rev. Mod. Phys.}\ }\textbf {\bibinfo {volume}
  {73}},\ \bibinfo {pages} {783} (\bibinfo {year} {2001})}\BibitemShut
  {NoStop}%
\bibitem [{\citenamefont {Bernardini}\ \emph {et~al.}(1997)\citenamefont
  {Bernardini}, \citenamefont {Fiorentini},\ and\ \citenamefont
  {Vanderbilt}}]{Bernardini1997}%
  \BibitemOpen
  \bibfield  {author} {\bibinfo {author} {\bibfnamefont {F.}~\bibnamefont
  {Bernardini}}, \bibinfo {author} {\bibfnamefont {V.}~\bibnamefont
  {Fiorentini}}, \ and\ \bibinfo {author} {\bibfnamefont {D.}~\bibnamefont
  {Vanderbilt}},\ }\href {\doibase 10.1103/PhysRevB.56.R10024} {\bibfield
  {journal} {\bibinfo  {journal} {Phys. Rev. B}\ }\textbf {\bibinfo {volume}
  {56}},\ \bibinfo {pages} {R10024} (\bibinfo {year} {1997})}\BibitemShut
  {NoStop}%
\bibitem [{\citenamefont {Wu}\ \emph {et~al.}(2012)\citenamefont {Wu},
  \citenamefont {Shivaraman}, \citenamefont {Wang},\ and\ \citenamefont
  {Speck}}]{Wu2012}%
  \BibitemOpen
  \bibfield  {author} {\bibinfo {author} {\bibfnamefont {Y.-R.}\ \bibnamefont
  {Wu}}, \bibinfo {author} {\bibfnamefont {R.}~\bibnamefont {Shivaraman}},
  \bibinfo {author} {\bibfnamefont {K.-C.}\ \bibnamefont {Wang}}, \ and\
  \bibinfo {author} {\bibfnamefont {J.~S.}\ \bibnamefont {Speck}},\ }\href
  {http://scitation.aip.org/content/aip/journal/apl/101/8/10.1063/1.4747532}
  {\bibfield  {journal} {\bibinfo  {journal} {Applied Physics Letters}\
  }\textbf {\bibinfo {volume} {101}},\ \bibinfo {eid} {083505} (\bibinfo {year}
  {2012})}\BibitemShut {NoStop}%
\bibitem [{\citenamefont {Nakamura}(1998)}]{Nakamura956}%
  \BibitemOpen
  \bibfield  {author} {\bibinfo {author} {\bibfnamefont {S.}~\bibnamefont
  {Nakamura}},\ }\href {\doibase 10.1126/science.281.5379.956} {\bibfield
  {journal} {\bibinfo  {journal} {Science}\ }\textbf {\bibinfo {volume}
  {281}},\ \bibinfo {pages} {956} (\bibinfo {year} {1998})},\ \Eprint
  {http://arxiv.org/abs/http://science.sciencemag.org/content/281/5379/956.full.pdf}
  {http://science.sciencemag.org/content/281/5379/956.full.pdf} \BibitemShut
  {NoStop}%
\bibitem [{\citenamefont {Anderson}(1958)}]{Anderson1958}%
  \BibitemOpen
  \bibfield  {author} {\bibinfo {author} {\bibfnamefont {P.~W.}\ \bibnamefont
  {Anderson}},\ }\href@noop {} {\bibfield  {journal} {\bibinfo  {journal}
  {Physical Review}\ }\textbf {\bibinfo {volume} {109}},\ \bibinfo {pages}
  {1492} (\bibinfo {year} {1958})}\BibitemShut {NoStop}%
\bibitem [{\citenamefont {Lee}\ and\ \citenamefont
  {Ramakrishnan}(1985)}]{Lee1985}%
  \BibitemOpen
  \bibfield  {author} {\bibinfo {author} {\bibfnamefont {P.~A.}\ \bibnamefont
  {Lee}}\ and\ \bibinfo {author} {\bibfnamefont {T.~V.}\ \bibnamefont
  {Ramakrishnan}},\ }\href {\doibase 10.1103/RevModPhys.57.287} {\bibfield
  {journal} {\bibinfo  {journal} {Review of Modern Physics}\ }\textbf {\bibinfo
  {volume} {57}},\ \bibinfo {pages} {287} (\bibinfo {year} {1985})}\BibitemShut
  {NoStop}%
\bibitem [{\citenamefont {Evers}\ and\ \citenamefont
  {Mirlin}(2008)}]{Evers2008}%
  \BibitemOpen
  \bibfield  {author} {\bibinfo {author} {\bibfnamefont {F.}~\bibnamefont
  {Evers}}\ and\ \bibinfo {author} {\bibfnamefont {A.~D.}\ \bibnamefont
  {Mirlin}},\ }\href@noop {} {\bibfield  {journal} {\bibinfo  {journal}
  {Reviews of Modern Physics}\ }\textbf {\bibinfo {volume} {80}},\ \bibinfo
  {pages} {1355} (\bibinfo {year} {2008})}\BibitemShut {NoStop}%
\bibitem [{\citenamefont {Lagendijk}\ \emph {et~al.}(2009)\citenamefont
  {Lagendijk}, \citenamefont {van Tiggelen},\ and\ \citenamefont
  {Wiersma}}]{Lagendijk2009}%
  \BibitemOpen
  \bibfield  {author} {\bibinfo {author} {\bibfnamefont {A.}~\bibnamefont
  {Lagendijk}}, \bibinfo {author} {\bibfnamefont {B.}~\bibnamefont {van
  Tiggelen}}, \ and\ \bibinfo {author} {\bibfnamefont {D.~S.}\ \bibnamefont
  {Wiersma}},\ }\href@noop {} {\bibfield  {journal} {\bibinfo  {journal}
  {Physics Today}\ }\textbf {\bibinfo {volume} {62}},\ \bibinfo {pages} {24}
  (\bibinfo {year} {2009})}\BibitemShut {NoStop}%
\bibitem [{\citenamefont {Rosenbaum}\ \emph {et~al.}(1983)\citenamefont
  {Rosenbaum}, \citenamefont {Milligan}, \citenamefont {Paalanen},
  \citenamefont {Thomas}, \citenamefont {Bhatt},\ and\ \citenamefont
  {Lin}}]{Rosenbaum1983}%
  \BibitemOpen
  \bibfield  {author} {\bibinfo {author} {\bibfnamefont {T.~F.}\ \bibnamefont
  {Rosenbaum}}, \bibinfo {author} {\bibfnamefont {R.~F.}\ \bibnamefont
  {Milligan}}, \bibinfo {author} {\bibfnamefont {M.~A.}\ \bibnamefont
  {Paalanen}}, \bibinfo {author} {\bibfnamefont {G.~A.}\ \bibnamefont
  {Thomas}}, \bibinfo {author} {\bibfnamefont {R.~N.}\ \bibnamefont {Bhatt}}, \
  and\ \bibinfo {author} {\bibfnamefont {W.}~\bibnamefont {Lin}},\ }\href
  {\doibase 10.1103/PhysRevB.27.7509} {\bibfield  {journal} {\bibinfo
  {journal} {Physical Review B}\ }\textbf {\bibinfo {volume} {27}},\ \bibinfo
  {pages} {7509} (\bibinfo {year} {1983})}\BibitemShut {NoStop}%
\bibitem [{\citenamefont {Vollhardt}\ and\ \citenamefont
  {W\"olfle}(1982)}]{Vollhardt1982}%
  \BibitemOpen
  \bibfield  {author} {\bibinfo {author} {\bibfnamefont {D.}~\bibnamefont
  {Vollhardt}}\ and\ \bibinfo {author} {\bibfnamefont {P.}~\bibnamefont
  {W\"olfle}},\ }\href {\doibase 10.1103/PhysRevLett.48.699} {\bibfield
  {journal} {\bibinfo  {journal} {Physical Review Letters}\ }\textbf {\bibinfo
  {volume} {48}},\ \bibinfo {pages} {699} (\bibinfo {year} {1982})}\BibitemShut
  {NoStop}%
\bibitem [{\citenamefont {Ostrovsky}\ \emph {et~al.}(2006)\citenamefont
  {Ostrovsky}, \citenamefont {Gornyi},\ and\ \citenamefont
  {Mirlin}}]{Ostrovsky2006}%
  \BibitemOpen
  \bibfield  {author} {\bibinfo {author} {\bibfnamefont {P.~M.}\ \bibnamefont
  {Ostrovsky}}, \bibinfo {author} {\bibfnamefont {I.~V.}\ \bibnamefont
  {Gornyi}}, \ and\ \bibinfo {author} {\bibfnamefont {A.~D.}\ \bibnamefont
  {Mirlin}},\ }\href {\doibase 10.1103/PhysRevB.74.235443} {\bibfield
  {journal} {\bibinfo  {journal} {Physical Review B}\ }\textbf {\bibinfo
  {volume} {74}},\ \bibinfo {pages} {235443} (\bibinfo {year}
  {2006})}\BibitemShut {NoStop}%
\bibitem [{\citenamefont {Mello}\ \emph {et~al.}(1988)\citenamefont {Mello},
  \citenamefont {Akkermans},\ and\ \citenamefont {Shapiro}}]{Mello1988}%
  \BibitemOpen
  \bibfield  {author} {\bibinfo {author} {\bibfnamefont {P.~A.}\ \bibnamefont
  {Mello}}, \bibinfo {author} {\bibfnamefont {E.}~\bibnamefont {Akkermans}}, \
  and\ \bibinfo {author} {\bibfnamefont {B.}~\bibnamefont {Shapiro}},\ }\href
  {\doibase 10.1103/PhysRevLett.61.459} {\bibfield  {journal} {\bibinfo
  {journal} {Physical Review Letters}\ }\textbf {\bibinfo {volume} {61}},\
  \bibinfo {pages} {459} (\bibinfo {year} {1988})}\BibitemShut {NoStop}%
\bibitem [{\citenamefont {Hofstetter}\ and\ \citenamefont
  {Schreiber}(1993)}]{Hofstetter1993}%
  \BibitemOpen
  \bibfield  {author} {\bibinfo {author} {\bibfnamefont {E.}~\bibnamefont
  {Hofstetter}}\ and\ \bibinfo {author} {\bibfnamefont {M.}~\bibnamefont
  {Schreiber}},\ }\href {\doibase 10.1103/PhysRevB.48.16979} {\bibfield
  {journal} {\bibinfo  {journal} {Physical Review B}\ }\textbf {\bibinfo
  {volume} {48}},\ \bibinfo {pages} {16979} (\bibinfo {year}
  {1993})}\BibitemShut {NoStop}%
\bibitem [{\citenamefont {Kalitsov}\ \emph {et~al.}(2012)\citenamefont
  {Kalitsov}, \citenamefont {Chshiev},\ and\ \citenamefont
  {Velev}}]{Kalitsov2012}%
  \BibitemOpen
  \bibfield  {author} {\bibinfo {author} {\bibfnamefont {A.~V.}\ \bibnamefont
  {Kalitsov}}, \bibinfo {author} {\bibfnamefont {M.~G.}\ \bibnamefont
  {Chshiev}}, \ and\ \bibinfo {author} {\bibfnamefont {J.~P.}\ \bibnamefont
  {Velev}},\ }\href {\doibase 10.1103/PhysRevB.85.235111} {\bibfield  {journal}
  {\bibinfo  {journal} {Physical Review B}\ }\textbf {\bibinfo {volume} {85}},\
  \bibinfo {pages} {235111} (\bibinfo {year} {2012})}\BibitemShut {NoStop}%
\bibitem [{\citenamefont {Filoche}\ and\ \citenamefont
  {Mayboroda}(2012)}]{Filoche2012}%
  \BibitemOpen
  \bibfield  {author} {\bibinfo {author} {\bibfnamefont {M.}~\bibnamefont
  {Filoche}}\ and\ \bibinfo {author} {\bibfnamefont {S.}~\bibnamefont
  {Mayboroda}},\ }\href@noop {} {\bibfield  {journal} {\bibinfo  {journal}
  {Proceedings of the National Academy of Sciences of the USA}\ }\textbf
  {\bibinfo {volume} {109}},\ \bibinfo {pages} {14761} (\bibinfo {year}
  {2012})}\BibitemShut {NoStop}%
\bibitem [{\citenamefont {Arnold}\ \emph {et~al.}(2016)\citenamefont {Arnold},
  \citenamefont {David}, \citenamefont {Jerison}, \citenamefont {Mayboroda},\
  and\ \citenamefont {Filoche}}]{Arnold2016}%
  \BibitemOpen
  \bibfield  {author} {\bibinfo {author} {\bibfnamefont {D.~N.}\ \bibnamefont
  {Arnold}}, \bibinfo {author} {\bibfnamefont {G.}~\bibnamefont {David}},
  \bibinfo {author} {\bibfnamefont {D.}~\bibnamefont {Jerison}}, \bibinfo
  {author} {\bibfnamefont {S.}~\bibnamefont {Mayboroda}}, \ and\ \bibinfo
  {author} {\bibfnamefont {M.}~\bibnamefont {Filoche}},\ }\href@noop {}
  {\bibfield  {journal} {\bibinfo  {journal} {Physical Review Letters}\
  }\textbf {\bibinfo {volume} {116}},\ \bibinfo {pages} {056602} (\bibinfo
  {year} {2016})}\BibitemShut {NoStop}%
\bibitem [{\citenamefont {Piccardo}\ \emph {et~al.}(2017)\citenamefont
  {Piccardo}, \citenamefont {Li}, \citenamefont {Wu}, \citenamefont {Speck},
  \citenamefont {Bonef}, \citenamefont {Farrell}, \citenamefont {Filoche},
  \citenamefont {Martinelli}, \citenamefont {Peretti},\ and\ \citenamefont
  {Weisbuch}}]{Piccardo2017}%
  \BibitemOpen
  \bibfield  {author} {\bibinfo {author} {\bibfnamefont {M.}~\bibnamefont
  {Piccardo}}, \bibinfo {author} {\bibfnamefont {C.-K.}\ \bibnamefont {Li}},
  \bibinfo {author} {\bibfnamefont {Y.-R.}\ \bibnamefont {Wu}}, \bibinfo
  {author} {\bibfnamefont {J.~S.}\ \bibnamefont {Speck}}, \bibinfo {author}
  {\bibfnamefont {B.}~\bibnamefont {Bonef}}, \bibinfo {author} {\bibfnamefont
  {R.~M.}\ \bibnamefont {Farrell}}, \bibinfo {author} {\bibfnamefont
  {M.}~\bibnamefont {Filoche}}, \bibinfo {author} {\bibfnamefont
  {L.}~\bibnamefont {Martinelli}}, \bibinfo {author} {\bibfnamefont
  {J.}~\bibnamefont {Peretti}}, \ and\ \bibinfo {author} {\bibfnamefont
  {C.}~\bibnamefont {Weisbuch}},\ }\href@noop {} {\bibfield  {journal}
  {\bibinfo  {journal} {Physical Review B}\ } (\bibinfo {year}
  {2017})}\BibitemShut {NoStop}%
\bibitem [{\citenamefont {Li}\ \emph {et~al.}(2017)\citenamefont {Li},
  \citenamefont {Piccardo}, \citenamefont {Lu}, \citenamefont {Mayboroda},
  \citenamefont {Martinelli}, \citenamefont {Speck}, \citenamefont {Weisbuch},
  \citenamefont {Filoche},\ and\ \citenamefont {Wu}}]{Li2017}%
  \BibitemOpen
  \bibfield  {author} {\bibinfo {author} {\bibfnamefont {C.-K.}\ \bibnamefont
  {Li}}, \bibinfo {author} {\bibfnamefont {M.}~\bibnamefont {Piccardo}},
  \bibinfo {author} {\bibfnamefont {L.-S.}\ \bibnamefont {Lu}}, \bibinfo
  {author} {\bibfnamefont {S.}~\bibnamefont {Mayboroda}}, \bibinfo {author}
  {\bibfnamefont {L.}~\bibnamefont {Martinelli}}, \bibinfo {author}
  {\bibfnamefont {J.~S.}\ \bibnamefont {Speck}}, \bibinfo {author}
  {\bibfnamefont {C.}~\bibnamefont {Weisbuch}}, \bibinfo {author}
  {\bibfnamefont {M.}~\bibnamefont {Filoche}}, \ and\ \bibinfo {author}
  {\bibfnamefont {Y.-R.}\ \bibnamefont {Wu}},\ }\href@noop {} {\bibfield
  {journal} {\bibinfo  {journal} {Physical Review B}\ } (\bibinfo {year}
  {2017})}\BibitemShut {NoStop}%
\bibitem [{\citenamefont {Agmon}(1982)}]{Agmon1982}%
  \BibitemOpen
  \bibfield  {author} {\bibinfo {author} {\bibfnamefont {S.}~\bibnamefont
  {Agmon}},\ }\href@noop {} {\emph {\bibinfo {title} {Lectures on exponential
  decay of solutions of second-order elliptic equations: bounds on
  eigenfunctions of N-body Schr\"odinger operators}}},\ \bibinfo {series}
  {Mathematical Notes}, Vol.~\bibinfo {volume} {29}\ (\bibinfo  {publisher}
  {Princeton University Press},\ \bibinfo {address} {Princeton, New Jersey},\
  \bibinfo {year} {1982})\BibitemShut {NoStop}%
\bibitem [{\citenamefont {Agmon}(1985)}]{Agmon1985}%
  \BibitemOpen
  \bibfield  {author} {\bibinfo {author} {\bibfnamefont {S.}~\bibnamefont
  {Agmon}},\ }\href@noop {} {\bibfield  {journal} {\bibinfo  {journal} {Lecture
  Notes in Mathematics}\ }\textbf {\bibinfo {volume} {1159}},\ \bibinfo {pages}
  {1} (\bibinfo {year} {1985})}\BibitemShut {NoStop}%
\bibitem [{\citenamefont {Bondar}\ and\ \citenamefont
  {Liu}(2011)}]{Bondar2011}%
  \BibitemOpen
  \bibfield  {author} {\bibinfo {author} {\bibfnamefont {D.~I.}\ \bibnamefont
  {Bondar}}\ and\ \bibinfo {author} {\bibfnamefont {W.-K.}\ \bibnamefont
  {Liu}},\ }\href@noop {} {\bibfield  {journal} {\bibinfo  {journal} {Journal
  of Physics A: Mathematical and Theoretical}\ }\textbf {\bibinfo {volume}
  {44}},\ \bibinfo {pages} {275301} (\bibinfo {year} {2011})}\BibitemShut
  {NoStop}%
\bibitem [{\citenamefont {Strauss}(2008)}]{Strauss2008}%
  \BibitemOpen
  \bibfield  {author} {\bibinfo {author} {\bibfnamefont {W.~A.}\ \bibnamefont
  {Strauss}},\ }\href@noop {} {\emph {\bibinfo {title} {Partial Differential
  Equations}}}\ (\bibinfo  {publisher} {John Wiley \& Sons},\ \bibinfo {year}
  {2008})\ Chap.~\bibinfo {chapter} {11}\BibitemShut {NoStop}%
\bibitem [{\citenamefont {van Roosbroeck}(1950)}]{vanRoosbroeck1950}%
  \BibitemOpen
  \bibfield  {author} {\bibinfo {author} {\bibfnamefont {W.}~\bibnamefont {van
  Roosbroeck}},\ }\href@noop {} {\bibfield  {journal} {\bibinfo  {journal}
  {Bell System Technical Journal}\ }\textbf {\bibinfo {volume} {29}},\ \bibinfo
  {pages} {560} (\bibinfo {year} {1950})}\BibitemShut {NoStop}%
\bibitem [{\citenamefont {J\"ungel}(2009)}]{Jungel2009}%
  \BibitemOpen
  \bibfield  {author} {\bibinfo {author} {\bibfnamefont {A.}~\bibnamefont
  {J\"ungel}},\ }\href@noop {} {\emph {\bibinfo {title} {Transport Equations
  for Semiconductors}}},\ \bibinfo {edition} {1st}\ ed.,\ \bibinfo {series}
  {Lecture Notes in Physics}, Vol.\ \bibinfo {volume} {773}\ (\bibinfo
  {publisher} {Springer-Verlag Berlin Heidelberg},\ \bibinfo {year}
  {2009})\BibitemShut {NoStop}%
\bibitem [{\citenamefont {Watson-Parris}\ \emph {et~al.}(2011)\citenamefont
  {Watson-Parris}, \citenamefont {Godfrey}, \citenamefont {Dawson},
  \citenamefont {Oliver}, \citenamefont {Galtrey}, \citenamefont {Kappers},\
  and\ \citenamefont {Humphreys}}]{WatsonParris2011}%
  \BibitemOpen
  \bibfield  {author} {\bibinfo {author} {\bibfnamefont {D.}~\bibnamefont
  {Watson-Parris}}, \bibinfo {author} {\bibfnamefont {M.~J.}\ \bibnamefont
  {Godfrey}}, \bibinfo {author} {\bibfnamefont {P.}~\bibnamefont {Dawson}},
  \bibinfo {author} {\bibfnamefont {R.~A.}\ \bibnamefont {Oliver}}, \bibinfo
  {author} {\bibfnamefont {M.~J.}\ \bibnamefont {Galtrey}}, \bibinfo {author}
  {\bibfnamefont {M.~J.}\ \bibnamefont {Kappers}}, \ and\ \bibinfo {author}
  {\bibfnamefont {C.~J.}\ \bibnamefont {Humphreys}},\ }\href@noop {} {\bibfield
   {journal} {\bibinfo  {journal} {Physical Review B}\ }\textbf {\bibinfo
  {volume} {83}},\ \bibinfo {pages} {115321} (\bibinfo {year}
  {2011})}\BibitemShut {NoStop}%
\bibitem [{\citenamefont {der Maur}(2015)}]{AufderMaur2015}%
  \BibitemOpen
  \bibfield  {author} {\bibinfo {author} {\bibfnamefont {M.~A.}\ \bibnamefont
  {der Maur}},\ }\href {\doibase 10.1007/s10825-015-0683-3} {\bibfield
  {journal} {\bibinfo  {journal} {Journal of Computational Electronics}\
  }\textbf {\bibinfo {volume} {14}},\ \bibinfo {pages} {398–408} (\bibinfo
  {year} {2015})}\BibitemShut {NoStop}%
\bibitem [{\citenamefont {Schulz}\ \emph {et~al.}(2015)\citenamefont {Schulz},
  \citenamefont {Caro}, \citenamefont {Coughlan},\ and\ \citenamefont
  {O'Reilly}}]{Schulz2015}%
  \BibitemOpen
  \bibfield  {author} {\bibinfo {author} {\bibfnamefont {S.}~\bibnamefont
  {Schulz}}, \bibinfo {author} {\bibfnamefont {M.~A.}\ \bibnamefont {Caro}},
  \bibinfo {author} {\bibfnamefont {C.}~\bibnamefont {Coughlan}}, \ and\
  \bibinfo {author} {\bibfnamefont {E.~P.}\ \bibnamefont {O'Reilly}},\ }\href
  {\doibase 10.1103/PhysRevB.91.035439} {\bibfield  {journal} {\bibinfo
  {journal} {Physical Review B}\ }\textbf {\bibinfo {volume} {91}},\ \bibinfo
  {pages} {035439} (\bibinfo {year} {2015})}\BibitemShut {NoStop}%
\bibitem [{\citenamefont {Singh}(2007)}]{Singh2007}%
  \BibitemOpen
  \bibfield  {author} {\bibinfo {author} {\bibfnamefont {J.}~\bibnamefont
  {Singh}},\ }\href@noop {} {\emph {\bibinfo {title} {Electronic and
  Optoelectronic Properties of Semiconductor Structures}}},\ \bibinfo {edition}
  {1st}\ ed.\ (\bibinfo  {publisher} {Cambridge University Press},\ \bibinfo
  {address} {New York, NY, USA},\ \bibinfo {year} {2007})\BibitemShut {NoStop}%
\bibitem [{\citenamefont {Han}\ \emph {et~al.}(2009)\citenamefont {Han},
  \citenamefont {Lee}, \citenamefont {Lee}, \citenamefont {Cho}, \citenamefont
  {Kwon}, \citenamefont {Lee}, \citenamefont {Noh}, \citenamefont {Kim},
  \citenamefont {Kim},\ and\ \citenamefont {Park}}]{Han2009}%
  \BibitemOpen
  \bibfield  {author} {\bibinfo {author} {\bibfnamefont {S.-H.}\ \bibnamefont
  {Han}}, \bibinfo {author} {\bibfnamefont {D.-Y.}\ \bibnamefont {Lee}},
  \bibinfo {author} {\bibfnamefont {S.-J.}\ \bibnamefont {Lee}}, \bibinfo
  {author} {\bibfnamefont {C.-Y.}\ \bibnamefont {Cho}}, \bibinfo {author}
  {\bibfnamefont {M.-K.}\ \bibnamefont {Kwon}}, \bibinfo {author}
  {\bibfnamefont {S.~P.}\ \bibnamefont {Lee}}, \bibinfo {author} {\bibfnamefont
  {D.~Y.}\ \bibnamefont {Noh}}, \bibinfo {author} {\bibfnamefont {D.~J.}\
  \bibnamefont {Kim}}, \bibinfo {author} {\bibfnamefont {Y.~C.}\ \bibnamefont
  {Kim}}, \ and\ \bibinfo {author} {\bibfnamefont {S.~J.}\ \bibnamefont
  {Park}},\ }\href@noop {} {\bibfield  {journal} {\bibinfo  {journal} {Applied
  Physics Letters}\ }\textbf {\bibinfo {volume} {94}},\ \bibinfo {pages}
  {231123} (\bibinfo {year} {2009})}\BibitemShut {NoStop}%
\bibitem [{\citenamefont {Song}\ \emph {et~al.}(2013)\citenamefont {Song},
  \citenamefont {Jeon}, \citenamefont {Hyoun~Joe}, \citenamefont {Kim},
  \citenamefont {Lee}, \citenamefont {Ah~Lee}, \citenamefont {Choi},
  \citenamefont {Sung}, \citenamefont {Kang}, \citenamefont {Choi},\ and\
  \citenamefont {Soo~Lee}}]{Song2013}%
  \BibitemOpen
  \bibfield  {author} {\bibinfo {author} {\bibfnamefont {H.}~\bibnamefont
  {Song}}, \bibinfo {author} {\bibfnamefont {K.-S.}\ \bibnamefont {Jeon}},
  \bibinfo {author} {\bibfnamefont {J.}~\bibnamefont {Hyoun~Joe}}, \bibinfo
  {author} {\bibfnamefont {S.}~\bibnamefont {Kim}}, \bibinfo {author}
  {\bibfnamefont {M.}~\bibnamefont {Lee}}, \bibinfo {author} {\bibfnamefont
  {E.}~\bibnamefont {Ah~Lee}}, \bibinfo {author} {\bibfnamefont
  {H.}~\bibnamefont {Choi}}, \bibinfo {author} {\bibfnamefont {J.}~\bibnamefont
  {Sung}}, \bibinfo {author} {\bibfnamefont {M.-G.}\ \bibnamefont {Kang}},
  \bibinfo {author} {\bibfnamefont {Y.-H.}\ \bibnamefont {Choi}}, \ and\
  \bibinfo {author} {\bibfnamefont {J.}~\bibnamefont {Soo~Lee}},\ }\href
  {\doibase http://dx.doi.org/10.1063/1.4823507} {\bibfield  {journal}
  {\bibinfo  {journal} {Applied Physics Letters}\ }\textbf {\bibinfo {volume}
  {103}},\ \bibinfo {eid} {141102} (\bibinfo {year} {2013}),\
  http://dx.doi.org/10.1063/1.4823507}\BibitemShut {NoStop}%
\bibitem [{\citenamefont {Kuo}\ \emph {et~al.}(2010)\citenamefont {Kuo},
  \citenamefont {Tsai}, \citenamefont {Yen}, \citenamefont {Hsu},\ and\
  \citenamefont {Shen}}]{Kuo2010}%
  \BibitemOpen
  \bibfield  {author} {\bibinfo {author} {\bibfnamefont {Y.~K.}\ \bibnamefont
  {Kuo}}, \bibinfo {author} {\bibfnamefont {M.~C.}\ \bibnamefont {Tsai}},
  \bibinfo {author} {\bibfnamefont {S.~H.}\ \bibnamefont {Yen}}, \bibinfo
  {author} {\bibfnamefont {T.~C.}\ \bibnamefont {Hsu}}, \ and\ \bibinfo
  {author} {\bibfnamefont {Y.~J.}\ \bibnamefont {Shen}},\ }\href {\doibase
  10.1109/JQE.2010.2045104} {\bibfield  {journal} {\bibinfo  {journal} {IEEE
  Journal of Quantum Electronics}\ }\textbf {\bibinfo {volume} {46}},\ \bibinfo
  {pages} {1214} (\bibinfo {year} {2010})}\BibitemShut {NoStop}%
\bibitem [{\citenamefont {Ancona}\ and\ \citenamefont
  {Tiersten}(1987)}]{Ancona1987}%
  \BibitemOpen
  \bibfield  {author} {\bibinfo {author} {\bibfnamefont {M.~G.}\ \bibnamefont
  {Ancona}}\ and\ \bibinfo {author} {\bibfnamefont {H.~F.}\ \bibnamefont
  {Tiersten}},\ }\href@noop {} {\bibfield  {journal} {\bibinfo  {journal}
  {Physical Review B}\ }\textbf {\bibinfo {volume} {35}},\ \bibinfo {pages}
  {7959} (\bibinfo {year} {1987})}\BibitemShut {NoStop}%
\bibitem [{\citenamefont {Ancona}\ and\ \citenamefont
  {Iafrate}(1989)}]{Ancona1989}%
  \BibitemOpen
  \bibfield  {author} {\bibinfo {author} {\bibfnamefont {M.~G.}\ \bibnamefont
  {Ancona}}\ and\ \bibinfo {author} {\bibfnamefont {G.~J.}\ \bibnamefont
  {Iafrate}},\ }\href@noop {} {\bibfield  {journal} {\bibinfo  {journal}
  {Physical Review B}\ }\textbf {\bibinfo {volume} {39}},\ \bibinfo {pages}
  {9536} (\bibinfo {year} {1989})}\BibitemShut {NoStop}%
\end{thebibliography}%

\end{document}